# Applications of Deep Learning Techniques for Automated Multiple Sclerosis Detection Using Magnetic Resonance Imaging: A Review

Afshin Shoeibi [1,*], Marjane Khodatars[2], Mahboobeh Jafari[3], Parisa Moridian[4], Mitra Rezaei[5], Roohallah Alizadehsani[6], Fahime Khozeimeh[6], Juan Manuel Gorriz[7,8], Jónathan Heras[9], Maryam Panahiazar[10], Saeid Nahavandi[6], U. Rajendra Acharya [11,12,13]

1 Faculty of Electrical Engineering, Biomedical Data Acquisition Lab (BDAL), K. N. Toosi University of Technology, Tehran, Iran.
2 Faculty of Engineering, Mashhad Branch, Islamic Azad University, Mashhad Iran.
3 Electrical and Computer Engineering Faculty, Semnan University, Semnan, Iran.
4 Faculty of Engineering, Science and Research Branch, Islamic Azad University, Tehran, Iran.
5 Electrical and Computer Engineering Dept., Tarbiat Modares University, Tehran, Iran.
6 Institute for Intelligent Systems Research and Innovation (IISRI), Deakin University, Geelong, Australia.
7 Department of Signal Theory, Networking and Communications, Universidad de Granada, Spain.
8 Department of Psychiatry. University of Cambridge, UK.
9 Department of Mathematics and Computer Science, University of La Rioja, La Rioja, Spain.
10 University of California San Francisco, San Francisco, CA, USA.
11 Department of Biomedical Engineering, School of Science and Technology, Singapore University of Social Sciences, Singapore.
12 Dept. of *Electronics* and Computer Engineering, Ngee Ann Polytechnic, Singapore 599489, Singapore.
13 Department of Bioinformatics and Medical Engineering, Asia University, Taiwan.

* Corresponding Author: afshin.shoeibi@gmail.com

**Abstract**
Multiple Sclerosis (MS) is a type of brain disease which causes visual, sensory, and motor problems for people with a detrimental effect on the functioning of the nervous system. In order to diagnose MS, multiple screening methods have been proposed so far; among them, magnetic resonance imaging (MRI) has received considerable attention among physicians. MRI modalities provide physicians with fundamental information about the structure and function of the brain, which is crucial for the rapid diagnosis of MS lesions. Diagnosing MS using MRI is time-consuming, tedious, and prone to manual errors. Research on the implementation of computer aided diagnosis system (CADS) based on artificial intelligence (AI) to diagnose MS involves conventional machine learning and deep learning (DL) methods. In conventional machine learning, feature extraction, feature selection, and classification steps are carried out by using trial and error; on the contrary, these steps in DL are based on deep layers whose values are automatically learn. In this paper, a complete review of automated MS diagnosis methods performed using DL techniques with MRI neuroimaging modalities is provided. Initially, the steps involved in various CADS proposed using MRI modalities and DL techniques for MS diagnosis are investigated. The important preprocessing techniques employed in various works are analyzed. Most of the published papers on MS diagnosis using MRI modalities and DL are presented. The most significant challenges facing and future direction of automated diagnosis of MS using MRI modalities and DL techniques are also provided.

**KeyWords:** Multiple Sclerosis, Diagnosis, MRI, Neuroimaging, Deep Learning

## 1. Introduction

Multiple sclerosis (MS) is a chronic autoimmune disease wherein the immune system wrongly targets the central nervous system including the brain and spinal cord [1]. In MS, the nervous system, including the myelin sheath, nerve fibers, and even the cells that produce the myelin, is usually damaged [2]. These injuries go away after few days to few weeks if they are not very severe, but may cause permanent changes in the spinal cord if they are severe [3, 4]. These permanent changes are called sclerosis, and because these lesions occur in multiple and different areas, the disease is called multiple sclerosis [5, 6]. Due to this disease, the body immune system responds abnormally, causing inflammation and damage to parts of the body [7-8].

Many people around the world suffer from this disease. It is estimated worldwide that the number of people suffering from this disease has increased to 2.3 million from 2013 [9-10]. Figure (1) shows the number of people with MS worldwide [11]. As shown in Figure (1), North America, Europe, and Australia have most of the MS patients [11].

There are four categories in MS: (i) clinically isolated syndrome (CIS) [12-13], (ii) relapsing-remitting MS (RRMS) [14-15], (iii) primary progressive MS (PPMS) [16-17], and (iv) secondary progressive MS (SPMS) [18, 19]. The CIS refers to the first episode of neurological symptoms that lasts at least 24 hours and is caused by inflammation or demyelination of the central nervous system (CNS) [12-13] [20]. CIS can be either mono-focal or multi-focal [12-13] [20]. The RRMS is the most common type of MS disease, characterized by clearly defined attacks, also known as relapses or exacerbations of new or growing neurological symptoms, with intervals of remission in between [14-15], [21]. During remission, all symptoms may disappear, or some of them may continue and become permanent. However, there is no apparent progression of the disease during these periods. RRMS can be active or inactive, worsening or not worsening [14-15], [21-22]. Around 85% of people with MS are initially diagnosed with RRMS type of disease [14-15], [21-22]. PPMS type of MS disease is characterized by worsening of neurological function (i.e., increased disability) since the onset of symptoms, with no early recurrence or recovery [1617], [23]. PPMS can be described as active or inactive, with or without progression [16-17], [23]. SPMS is followed by an early relapse period. Few people with RRMS eventually switch to SPMS, wherein there is a gradual deterioration of nerve function over time [18-19], [24]. It can be also be characterized as either active or inactive, with or without progression [18-19], [24].

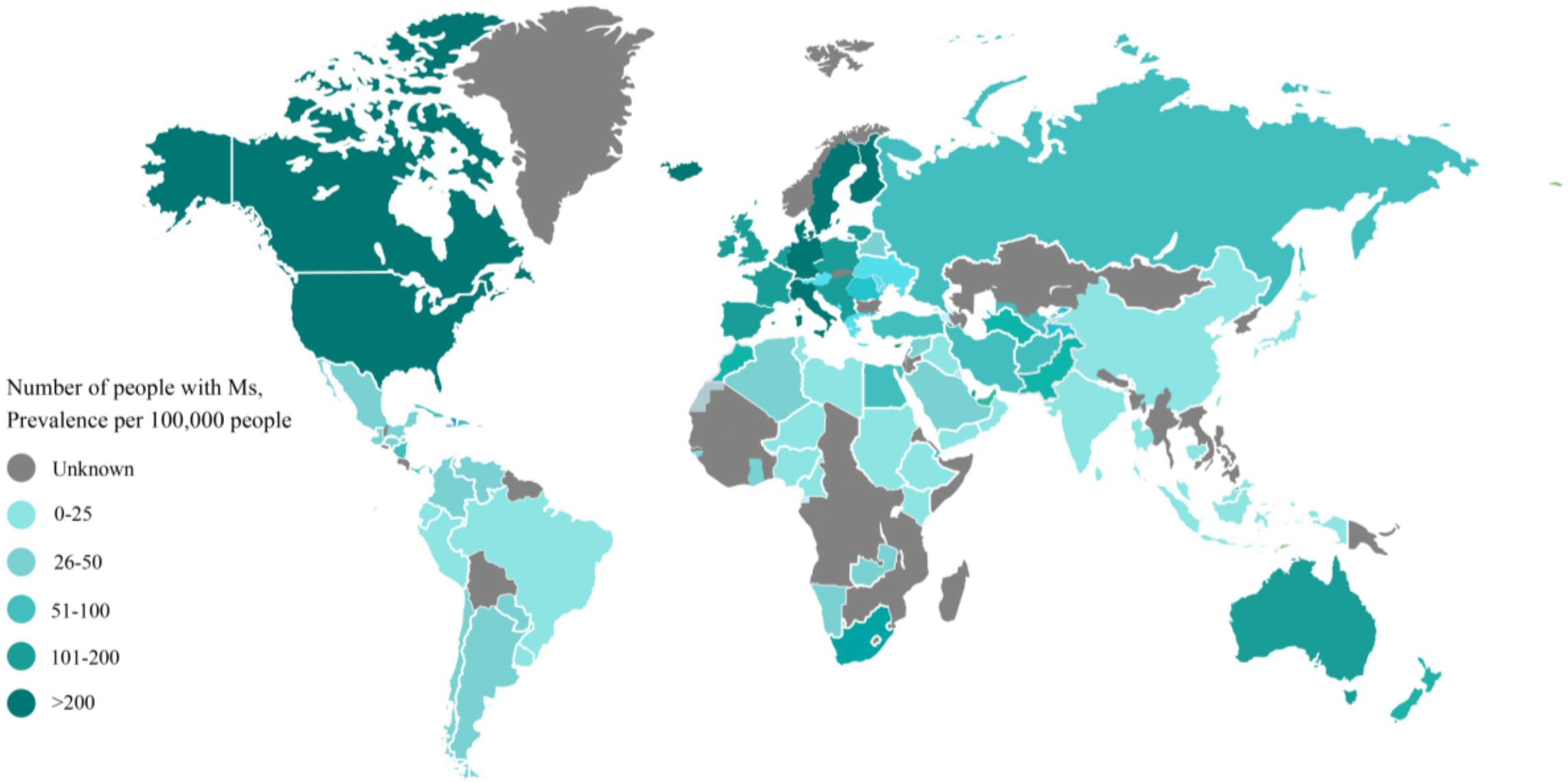

.Fig. 1. Number of MS patients worldwide

Common symptoms of MS include fatigue, difficulty in walking, spasticity, weakness, vision problems, dizziness, cognitive changes, emotional changes, depression, and more [25-27]. There are no known causes of MS disease [28-29]. Scientists believe that a combination of environmental and genetic factors play a role in MS [30-31]. Environmental factors such as geography, vitamin D deficiency, obesity, and smoking may have some correlation with MS [32-33].

Unfortunately, there are currently no symptoms, physical findings, or laboratory tests to accurately diagnose MS [34-35]. For this reason, several methods are used to diagnose MS that include reviewing the patient's medical history [36-37], medical imaging techniques such as MRI [38-40], spinal fluid analysis [41-42], and blood tests [43-44]. Currently, MRI modalities are the best non-invasive method used for the diagnosis of MS [45-47]. The myelin sheath, which protects nerve cell fibers, contains fat and repels water. In areas where MS damages myelin sheath, fat is stripped away. As fat is lost, this area keeps more water, and depending on the type of MRI scan, it is seen as a light white spot or lesions [48-49]. Figure (2) displayed different structural MRI (sMRI) modalities for different subjects with the MS disease [79-80].

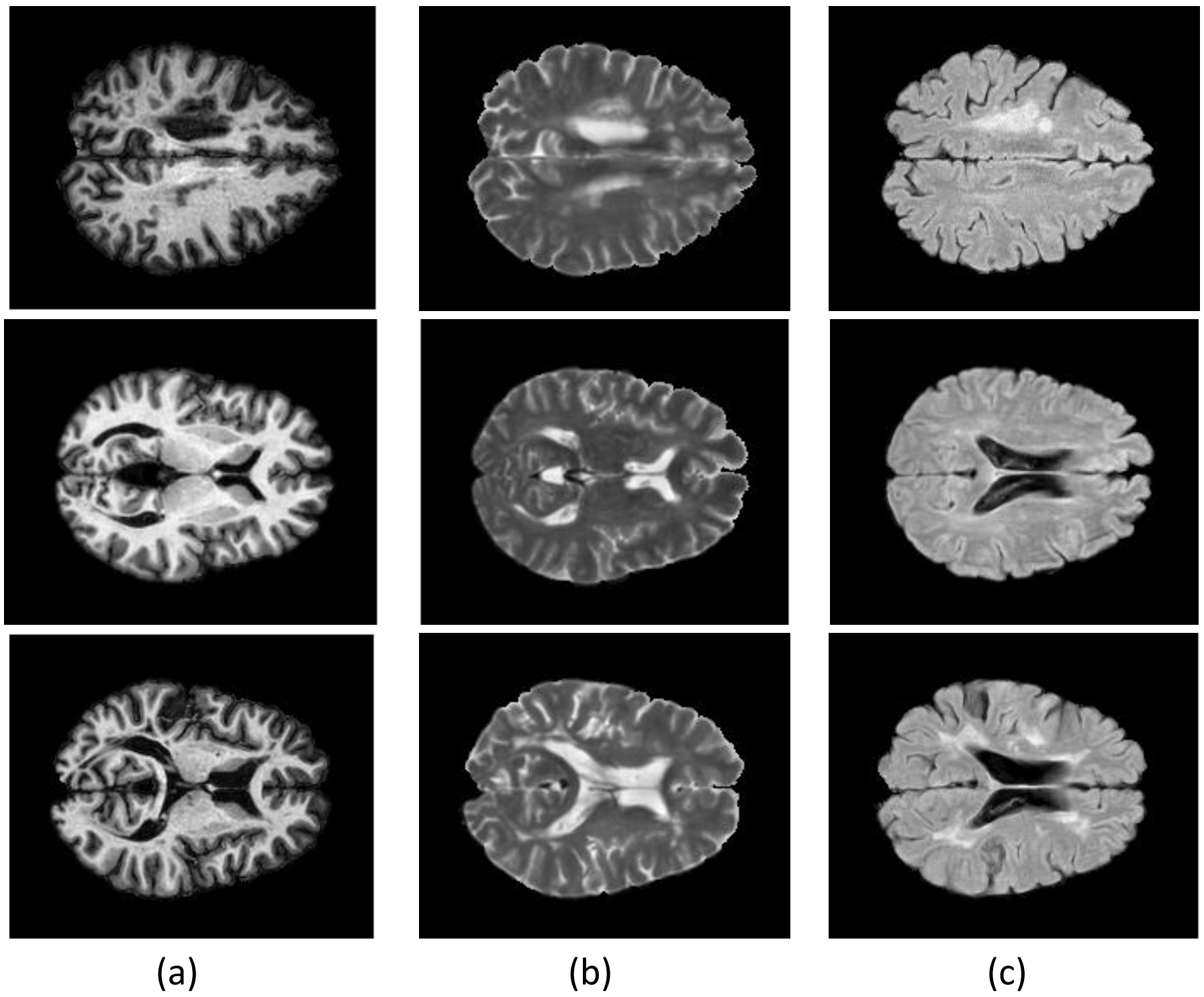

Fig. 2. Diagnosis of MS for different subjects using sMRI neuroimaging modalities [79-80].

Figure (2), illustrates various sMRI modalities used for the diagnosis of MS which were recorded by a 3T scanner [79-80]. Figure (2a) shows the T1-weighted images of MS patients. The T2-weighted images of different MS patients are displayed in Figure (2b). Figure (2c) shows the T2-weighted FLAIR images of MS patients.

Diagnosis of MS based on MRI neuroimaging modalities are time-consuming and challenging for physicians. Therefore, researchers are proposing novel methods to accurately identify these fields. Nowadays, AI techniques have emerged as an important tool for various disease diagnoses with the help of physicians [50-54]. AI techniques for medical diagnosis can be mainly split into (i) conventional machine learning methods and (ii) DL techniques [55-58].

In preliminary research, the conventional machine learning methods were used for diagnosis of MS using MRI modalities [27-29]. Use of these methods for the diagnosis of MS are highly time-consuming and requires significant amount of expertise in various AI fields. Also, several deficiencies related to conventional machine learning CADS include high computational cost due to multiple algorithms, inefficient performance with large amount of MRI input data, and so on.

On the other hand, the DL methods are one of the latest fields of AI which has gained considerable popularity to diagnose a variety of diseases using medical data [59-61]. One of the prominent features of DL networks is their ability to infer and derive inherent and latent feature representations in MRI data [55]. Another advantage of DL is that it does not require any manual management of the feature extraction stage; this allows the integration of feature extraction and classification steps in CADS using DL [90-93]. Research in the field of MS diagnosis using MRI modalities and DL architectures has been initiated since 2016. The research in this field comprised of the utilization of DL models for segmentation and classification applications.

This paper provides a review of studies conducted on the diagnosis of MS using MRI modalities and DL techniques. DL techniques are used to segment the MS lesions, and to detect MS automatically using MRI modalities. These techniques are discussed in detail in the following sections.

Figure (3) shows the number of papers published in various journals on MS diagnosis using MRI modalities and DL techniques. The keywords "MS", "MRI", "sMRI", "Multiple sclerosis", and "Deep Learning" have been used to search the papers in various scientific databases. Google Scholar has also been used for search. It can be noted from the Figure that research into the diagnosis of MS using MRI modalities and DL techniques started in 2016, and most of the papers are published in IEEE journals.

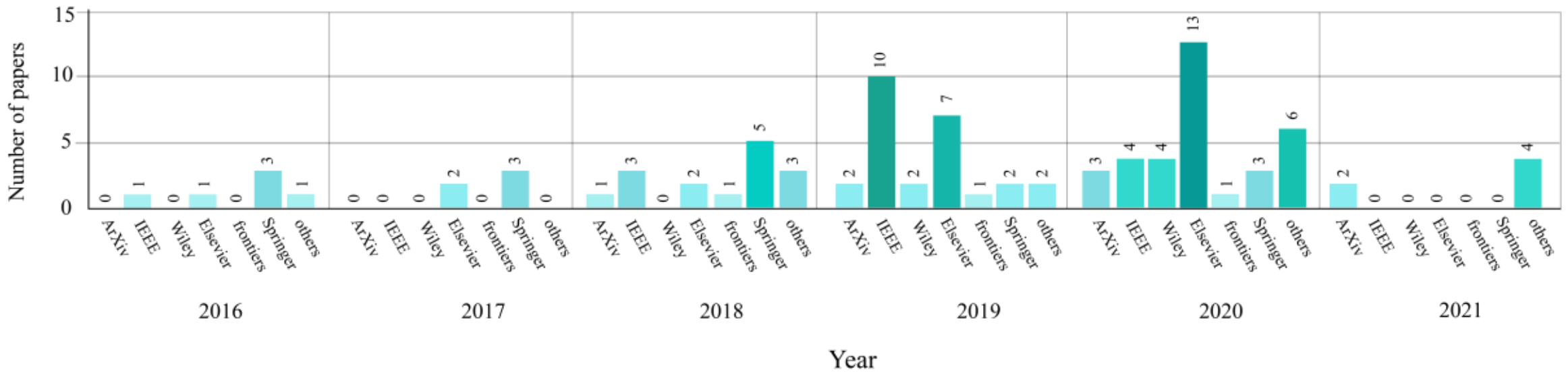

Fig. 3. Number of papers published on MS diagnosis yearly.

The organization of the rest of the paper is as follows. In the second section, CADS for MS detection using MRI neuroimaging modalities and DL networks are described. Discussion on the various works done is provided in Section 3. The challenges in the accurate diagnosis of MS using MRI modalities and DL are presented in Section 4. In section 5, future directions for the automated MS diagnosis using DL techniques are expressed. Finally, conclusions are presented in Section 6.

## 2. CADS for MS detection in MRI modalities

Today, CADS based on AI techniques are exploited in various medical applications [62-65]. The CADS in medicine can be implemented using conventional machine learning and DL methods [66-68]. Generally, a CADS construction comprises of dataset, preprocessing, feature extraction, feature selection, classification, and model evaluation steps [69-70]. One of the significant applications of CADS is the diagnosis of MS using MRI neuroimaging modalities. So far, numerous works have been accomplished on the implementation of CADS based on conventional machine learning [27-29] and DL [168-170] techniques for MS detection.

The main difference in CADS based on conventional machine learning and DL is in the feature extraction and feature selection steps [71-72]. CADS utilizing conventional machine learning involve feature extraction and feature selection algorithms using trial and error methods, requiring prior knowledge of image processing and AI techniques [73-74]. Low performance is obtained with large amount of MRI is another limitation of conventional machine learning-based CADS. Another disadvantage of these methods in dealing with high artifacts and low contrast of MR images.

However, in CADS based on DL, these two steps are performed intelligently by deep layers [75-76]. One advantage of DL is augmentation of input data does not diminish the performance of CADS [55]. Additionally, DL models are robust to noisy MRI data [55]. Therefore, in conclusion, it can be concluded that DL-based CADS have yielded higher performances in detecting MS using MRI modalities compared to conventional machine learning. Figure (4) shows the CADS block diagram for MS detection using DL techniques and MRI modalities. First, the datasets available are fed to the MS diagnosis system. Then, a variety preprocessing techniques are performed on MRI modalities. Finally, a robust and accurate DL architecture is obtained to detect the MS automatically.

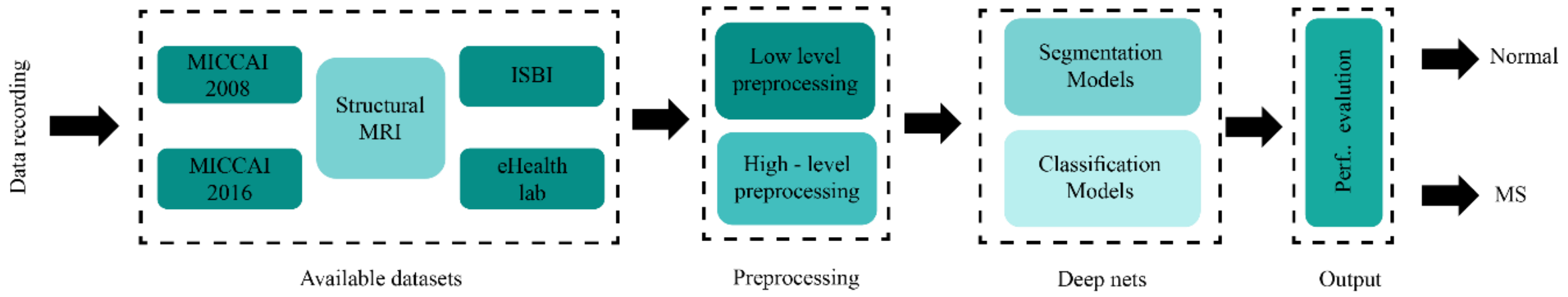

Fig. 4. Block diagram of CAD system using DL architecture for MS detection.

## 2.1. Datasets

In this section, the most important available datasets used to diagnose MS using MRI neuroimaging modalities are discussed. There are several datasets available for researchers to diagnose MS, including MICCAI 2008 [77], MICCAI 2016 [78], ISBI 2015 [79-80], and eHealth Lab [81]. In the following, the details for available MS datasets based on MRI neuroimaging modalities are presented. Also, a summary information of these datasets are provided in Table (1).

### 2.1.1. The MICCAI 2008 MS Lesion Segmentation Challenge Dataset

This dataset contains MR images of research subjects from the University of North Carolina (UNC) and Boston Children's Hospital (CHB) [77]. All dataset images were segmented by one CHB expert and 2 UNC experts. The training database has 20 samples, of which 10 CHB samples and 10 UNC samples were manually segmented from the CHB expert [77]. The test dataset includes 25 samples (15 CHB samples and 10 UNC samples) without any segmentation [77]. In the dataset, all the modalities of T1WI, T2WI, FLAIR, DTI-derived FA, and MD images are presented. Also, several pre-processing steps have been performed on this database. More information is provided in [77].

### 2.1.2. The MICCAI 2016 MS Lesion Segmentation Challenge Dataset

The 2016 MICCAI dataset includes MR images of 53 people with MS [78]. The images in this dataset were recorded from three different centers in France with four MRI scanners (Siemens, Philips, and GE) and include three 3T magnets and one 1.5T magnet [78]. In this dataset, MR images have also been manually segmented by seven experts [78]. This dataset is divided into training and testing. The training and testing set includes 15 and 38 patients, respectively [78]. Modalities of 3D FLAIR sequence, T1 weighted sequence pre, and post-Gadolinium injection, axial dual PD-T2 weighted sequence are provided for each patient. This challenge provided raw and preprocessed data for each patient [78].

### 2.1.3. ISBI 2015 Longitudinal MS Lesion Segmentation Challenge Dataset

The ISBI 2015 dataset includes MR images of 19 MS patients with a training and two test packages [79-80]. The training set has five subjects, four subjects with four-time points and one subject with five-time points [79-80]. The test set A consists of 10 subjects, eight of them with four-time points, one with five-time points, and one with six-time points [79-80]. Test set B has four subjects, three with four-time points and two with five-time points. All subjects have T1-w MPRAGE, T2-w & PD-w DSE, and T2-w FLAIR modalities. In this dataset, the original images, as well as the pre-processed images, are available. Also, manual segmentation has been conducted by two experts [79-80].

### 2.1.4. eHealth Lab

This dataset provides MRI modalities of 38 patients (17 males, 21 females) with a mean age of 34.1 ± 10.5 years with CIS of MS and MRI brain lesions, recorded twice with an interval of 6-12 months and with 1.5 T protocol [81].

Table 1. Details of public datasets available for MS diagnosis

| Ref | Dataset | Number of Cases | | Modalities | Description | Link |
|---|---|---|---|---|---|---|
| [77] | MICCAI 2008 | Train | 20 | T1WI, T2WI, FLAIR, DTI-derived FA and MD | Preprocessed | https://www.nitrc.org/projects/msseg |
| | | Test | 25 | | | |
| [78] | MICCAI 2016 | Train | 15 | T1-w weighted, T1-w gadolinium enhanced (T1-w Gd), T2-w, T2-FLAIR and PD-w images | Preprocessed | http://www.miccai2016.org/en/ |
| | | Test | 38 | | | |
| [79-80] | ISBI 2015 | Train | 5 | T1-w MPRAGE, T2-w & PD-w DSE, T2-w FLAIR | Preprocessed | https://smart-stats-tools.org/lesion-challenge-2015 |
| | | Test A | 10 | | | |
| | | Test B | 4 | | | |
| [81] | eHealth Lab | 38 | | MRI | -- | http://www.medinfo.cs.ucy.ac.cy/index.php/facilities/32-software/218-datasets |

### 2.2. Preprocessing

Diagnosis of brain lesions using MRI modalities are clinically important for the diagnosis of MS. Segmentation and classification of brain lesions from MRI modalities are extremely problematic for physicians and are prone to misdiagnosis. Various factors such as artifacts, intensity heterogeneity, etc. have a destructive effect on the quality of MR image, which often can lead to misdiagnosis of the disease. In the following, the low level and high level preprocessing methods in MRI neuroimaging modalities for diagnosis of MS are discussed. Expressing these items prevents additional explanations of common preprocessing methods in Table (2).

#### 2.2.1. Low level preprocessing

In this section, the most important low level preprocessing methods for sMRI modalities are introduced, which include denoising, inhomogeneity correction, skull-stripping, registration, intensity standardization, de-oblique, re-orientation and segmentation. These preprocessing methods are consistent for all sMRI modalities.

**(1). Denoising**

During the MRI recording process, images are usually corrupted by various random noises [82]. Hence, several approaches are utilized to remove noise from MRI modalities. Some of those methods are low-pass filters [83], Fourier filters [84], and wavelets [85].

**(2). Inhomogeneity correction**

When the magnetic fields of MRI scanner strike the brain tissue, their intensity decline, creating an artifact in the images [82] [86]. This artifact is observed as a low-frequency variation in signal intensity of MRI images and should therefore be modified in the preprocessing step [82] [86]. Two important methods are applied for inhomogeneity correction [82] [86]. The first category is the expectation-maximization (EM) algorithm that models the bias field during the segmentation process [87] and the second category uses image properties [88].

**(3). Non-brain Tissue Removal (skull-stripping)**

In neuroimaging studies, the regions of interest (ROI) are located in the brain tissue. Therefore, non-brain tissues such as skull, neck, eyes, nose, and mouth are not important and should be eliminated [82] [86]. This enhances the accuracy of CADS for MS detection [82] [86].

**(4). Registration**

In MRI preprocessing, image registration is a prevalent step used to combine different types of image modalities or sequences (T1- and T2-weighted images of the same subject) or to place images in a standard space such as MNI [82] [86].

**(5). Intensity standardization**

Typically, MR images acquired with the same protocol do not contain the same intensity among scanners. Even in a scanner with the same settings, in various sessions, the image intensity patterns vary [82] [86]. Intensity standardization techniques in MRI modalities attempt to correct these scanner-dependent intensity variations [82] [86]. The most popular procedures used for intensity standardization in MRI modalities are histogram matching techniques [82] [86].

**(6). De-Oblique**

The oblique scanning is used to cover the whole brain while avoiding the artifacts caused by air and humidity in the eyes and nose. However, the oblique scanning makes registration of two different MR images challenging. So, a de-oblique preprocessing step should be done before registration [82] [86].

**(7). Re-orientation**

The direction of the image depends on the settings of image registration process. Differences in direction may lead to misregistration, so all images must have identical directions. Hence, re-orientation techniques are employed [82] [86].

**(8). Segmentation**

The aim of segmentation is to map the image into a set of meaningful areas containing identical characteristics in terms of intensity, depth, color, or structure [82] [86] [89]. In MRI modalities, the purpose of segmentation is to isolate three types of tissues: white matter (WM), gray matter (GM), and cerebrospinal fluid (CSF) [89].

#### 2.2.2. High level preprocessing (Others Preprocessing)

High level preprocessing techniques along with low level preprocessing methods help to improve the performance of CADS. These methods include data augmentation (DA) [135], patch extraction [168], ROIs extraction [182], and so on. Detailed preprocessing information employed by each group (paper) for the diagnosis of MS using DL methods and MRI modalities are summarized in Table (2).

### 2.3. Deep Leaning Methods used for MS detection

Nowadays, DL techniques are used in various medical fields attracting lot of attention from many researchers [90-93]. One of these areas is the diagnosis of brain diseases such as MS using MRI modalities. Table (2) shows the types of DL networks used in MS diagnostic research using MRI modalities. It can be noted from after table (2), the most popular DL architectures used for MS detection use convolutional neural networks (CNNs) [94-96], Autoencoders (AEs) [94-96], generative adversarial networks (GANs) [97-99], and CNN-RNN models [100]. CNNs are the first class of DL techniques used in the supervised learning methods category. It can be noted from Table (2), that most of the researchers used various 2D-CNN and 3D-CNN models for segmentation and classification of MRI modalities for MS detection. AEs are a group of DL networks that are based on unsupervised learning. GAN architectures are the recently developed DL models used for the MS diagnosis using MRI modalities. In the following sections, we have briefly discussed these various DL networks that have been used for the diagnosis of MS. Also, details of DL networks are provided in the appendix table A.

#### 2.3.1. Convolutional Neural Networks

An advantage of CNN models is that they do not need manual feature extraction. In these models, as the network becomes deeper, the higher level features are extracted [94-96]. Although CNNs have exhibited an acceptable performance, this performance promotion has been attained at the expense of increased computational complexities and, therefore, there is a need for more powerful processing hardware such as Graphical Processing Unit (GPU) and Tensor Processing Unit (TPU) [94-96]. In this section, the most significant CNN architectures used for automatic segmentation and classification of MS using MRI neuroimaging modalities are presented. First, different 2D and 3D CNNs models developed for MS classification are presented. Then, CNN architectures namely U-Net [101] and FCN [102] developed for MR images segmentation are discussed.

**(1). 2D and 3D-CNNs**

CNNs are one of the most popular DL techniques, with a variety of applications, including image segmentation [103-104], image classification [105, 106], and more [107-108]. These networks have a better compatibility with 2D and 3D images due to the reduction of parameter numbers and ability to reuse weights [94-96]. The most important components of a CNN architecture are convolutional, pooling, batch normalization, and fully connected layers [94-96]. By properly selecting layers on CNN, it is possible to learn the spatial and temporal dependencies of an image [94-96]. Figure (5) shows a 2D-CNN architecture for classification of MR images.

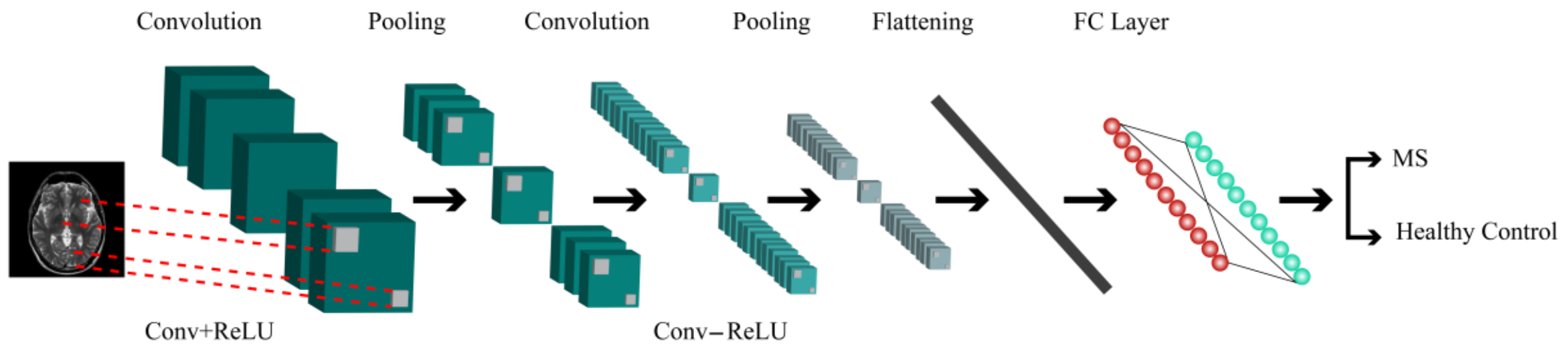

Fig. 5. Block diagram of 2D-CNN used for automated MS detection.

**(2). Pre-Trained CNN Networks**

DL architectures have numerous training parameters; hence, when small datasets are used to train DL networks from scratch, they do not yield good classification results. Therefore, transfer learning with pre-trained models can be used to address these issues. Pre-trained models are first trained on a large dataset, such as ImageNet, and then their classification layer is replaced with a new layer specific for the problem at hand [110-111]. Subsequently, by feeding new data as input to the pre-trained models, their weights are updated, enabling them to classify the data [112]. The disadvantage of these models is the use of ImageNet data for initial training, while MRI images are of gray-scale type. The most popular pre-trained models for classifying MRI modalities for MS diagnosis include LeNet, AlexNet, GoogleNet, VGGNet, ResNet, etc. [109-112].

**(3). FCN Network**

This network was introduced by Long et al., which has taken advantage of available CNNs that learn hierarchies of features [102]. In this model, popular networks have transformed entirely convolutional models by replacing FC layers with convolution layers to capture output as a local map. These maps are up-sampled using the introduced method. The deconvolution method is as follows: to simulate up-sampling with size f, backward convolution method with stride size f is employed on the output. These layers are also capable of learning. At the end of the network, there is a 1x1 convolution layer that yields the corresponding pixel label as the output. The exiting stride in the deconvolution stage constraints the output detail quantity of this layer. To address this issue and enhance the quality of results, several skip connections have been added to the network from the lower layers to the end layer [102]. The main advantage of FCN is that it receives the input data with an arbitrary size, and produces a corresponding-sized output with efficient inference and learning [102]. The upsampling results are relatively fuzzy and insensitive to image details; segmentation results are not good enough, which is the main disadvantage of this network. Figure (6) shows the general fully convolutional network (FCN) block diagram used for automated brain MR images segmentation.

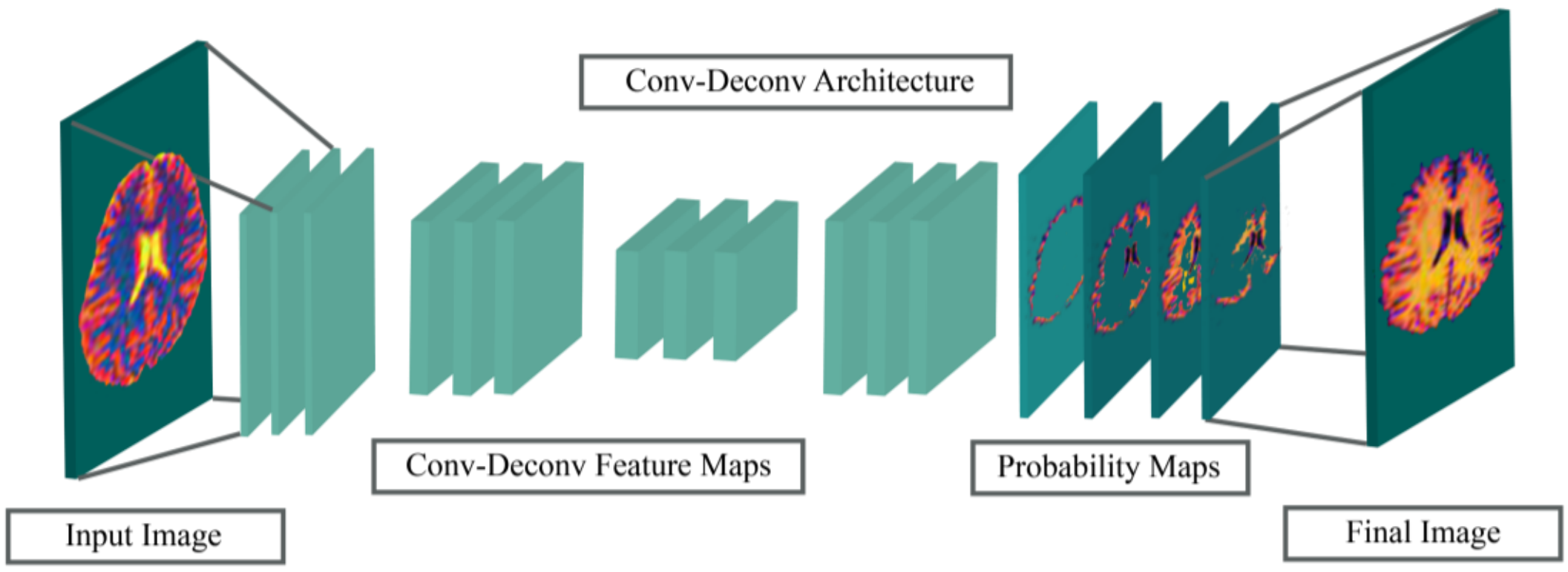

Fig. 6. Block diagram of FCN used for automated MS detection

**(4). U-Net**

U-Net is a well-known CNN architecture used for image segmentation that was first introduced by Ronneberger et al. [101]. This network possesses two parts: an encoder and a decoder, by which image segmentation operations are carried out [101]. In U-Net, the encoder section consists of several down-sampling and convolutional layers [101]. The decoder section also comprises a number of up-sampling and convolution layers. In this network, skip connections relations are placed between the corresponding up-sampling and down-sampling layers [101]. An advantage of U-Net models is that, they can be trained with a limited number of data. The main difference between U-Net and FCN is that the former is symmetrical and uses skip connections between two upsampling and down-sampling paths [101]. Figure (7) shows the general U-Net block diagram used for brain MR images segmentation.

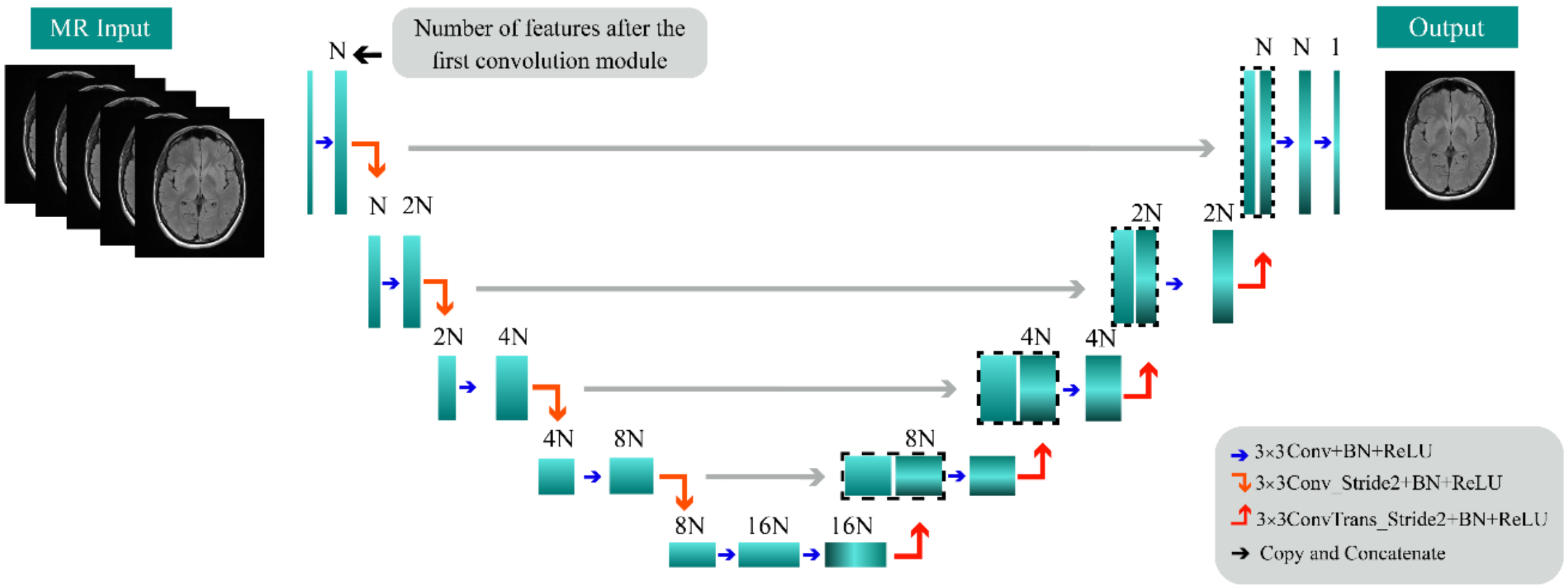

Fig. 7. Block diagram of U-Net used for automated MS detection.

**(5). Generative adversarial network (GAN)**

GAN architectures are a novel class of DL models applied to a wide variety of applications in various fields [98-99]. In general, GAN networks consist of two neural networks called the generator, G, and the discriminator, D, [98-99]. The role of generator is to estimate the probability distribution of the original data to generate samples similar to the original data [98-99]. The Discriminator, on the other hand, is trained to determine by likelihood estimation whether the sample is from original data or artificial data generated by the generator [98-99]. The term GAN is used because the generator and discriminator are trained to compete with each other. In this way, the generator tries to mislead the discriminator, whereas the discriminator attempts to identify better-generated samples [98-99]. The advantage of GAN models is that they do not require prior assumptions about the dataset, and are aimed to function with all data distributions [98-99]. The limitation of these models are the gradient vanishing problem and training complexity, which can be resolved to some extent by performing mathematical calculations in the network training phase. Figure (8) displays a GAN model used for the brain MR images classification.

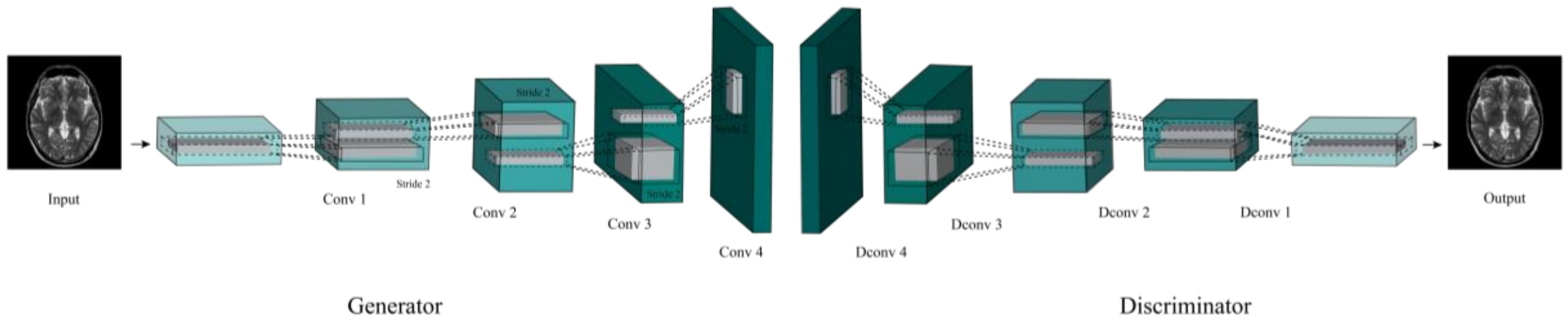

Fig. 8. Block diagram of GAN used for automated MS detection.

**2.3.2. Autoencoder**

Autoencoders (AEs) are a particular type of DL network aimed to find a low-dimensional representation of input data [95-96]. These models consist of two parts, an encoder and a decoder. The encoder compresses high-dimensional input data into lower-dimensional displays, known as latent space or bottleneck representation [95, 96]. The decoder returns the data to the original dimensions of the input. Denoising AE, Sparse AE, and Stacked AE are the most significant types of AE [95, 96].

**2.3.3. CNN-RNN**

Hybrid CNN-RNN architecture have become popular among AI professionals. This is due to the ability of CNN networks to learn spatial features and the ability of Recurrent Neural Network (RNN) architectures to learn temporal features [100]. In CNN-RNN architectures, data is often first fed to the CNN network input, and after passing through several layers of convolution, feature maps that are the output of the CNN network are applied to an RNN network [100]. The results reveal that adopting hybrid models such as CNN-RNN has been extremely successful in increasing the accuracy of CADS in brain disease diagnosis. Figure (9) shows the CNN-RNN architecture for MR images classification.

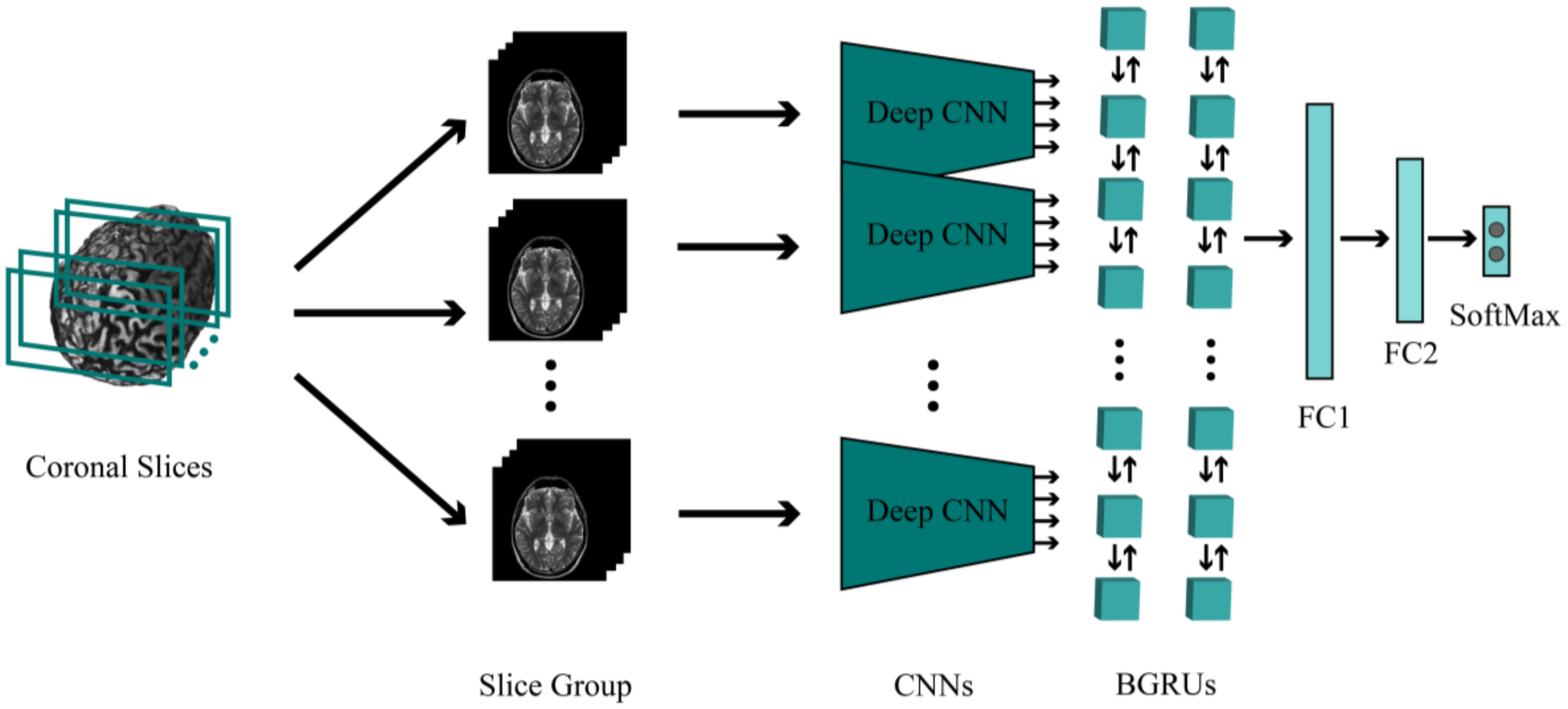

Fig. 9. Block diagram of CNN-RNN used for automated MS detection.

Table 2. Summary of CADS developed for MS using MRI neuroimaging modalities and DL methods.

| Works | Application | Dataset | Modalities | Number of Cases | Preprocessing Toolbox | Others Preprocessing | DNN | Toolbox | K Fold | Performance Criteria (%) |
|---|---|---|---|---|---|---|---|---|---|---|
| Marzullo [149] | EDSS Estimation | Clinical | MRI | 83 MS | -- | Brain Structural Connectivity Generation | 2D-CNN | -- | 5 | RMSE= 0.09 |
| Siar [154] | Diagnosing and Classification | Clinical | MRI | 320 MS, 791 HC | -- | -- | 2D-CNN | -- | -- | Acc=96. 88<br>Sen=94.64<br>Spec=100 |
| Aslani [155] | Segmentation | ISBI 2015 | MRI | 19 MS | -- | Data Augmentation (DA) | 2D-CNN | Keras | -- | DSC=69.80 |
| Eitel [158] | Harnessing Spatial MRI Normalization | Clinical | MRI | 76 MS, 71 HC | -- | DA | 2D-CNN | -- | -- | Acc=80.92 |
| Afzal [159] | Classification | John Hunter Hospital’s Dataset | MRI | 21 Patients | -- | DA | 2D-CNN | Keras | -- | Acc=100 |
| Roy [172] | Lesion Segmentation | ISBI 2015<br>Clinical | MRI | 19 MS<br>128 MS | -- | -- | 2D-CNN | TensorFlow, Keras | -- | DSC=52.4 |
| Aslani [184] | Lesion Segmentation | NRU Dataset | MRI | 37 MS | FSL | Decomposing 3D Data Into 2D Images | 2D-CNN | Keras, TensorFlow | 4 | DSC= 66.55 |
| Alijamaat [190] | Identification | eHealth Laboratory | MRI | 38 MS, 20 HC | -- | DA, Histogram Stretching, DWT | 2D-CNN | Keras, TensorFlow | -- | Acc=99.05<br>Sen=99.14<br>Spec=98.89<br>Prec=98.43 |
| Shrwan [223] | Classification | Clinical | MRI | 38 MS | -- | -- | 2D-CNN | MATLAB R2020a | -- | Acc=99.55<br>Pre=99.15<br>F1-S= 99.15 |
| Afzal [225] | Segmentation | MICCAI 2016<br>ISBI 2015 | MRI | 45 Scans<br>82 Scans | FSL | Patch Extraction | Two 2D-CNN | Keras, TensorFlow | -- | DSC=67<br>Sen=48<br>Pre=90 |
| Wang [181] | MS Identification | eHealth Laboratory and Private Data | MRI | 38 MS, 26 HC | -- | HS, DA | 2D-CNN | -- | -- | Acc=98.77<br>Sen=98.77<br>Spec=98.76 |
| Ulloa [135] | Segmentation | ISBI 2015 | MRI | 19 MS | -- | ICBM452 Atlas, Patch Extraction, DA | V-Net CNN | Keras, TensorFlow | 5 | DSC=68.77 |
| Birenbaum [179] | Lesion Segmentation | ISBI 2015 | MRI | 19 MS | -- | Lesion Extraction, DA | 4 CNN Models | Keras, Theano | 5 | Score=91.267<br>DSC=62.7 v |
| Birenbaum [165] | Lesion Segmentation | ISBI 2015 | MRI | 19 MS | -- | Candidate Extraction, DA | 4 CNN Models | Keras, Theano | 5 | DSC=62.7 |
| SALEM [156] | Generating Synthetic MS Lesions | Clinical | MRI | 65 MS, 15 HC | Nifty Reg Tools, ROBEX Tool, ITK Library | WMH Mask and the Intensity Level Masks, WMH FILLING, DA | Encoder-Decoder U-NET | Python, Keras, TensorFlow | -- | DSC=63<br>Sen=55<br>Pre=79 |
| | | ISBI 2015 | | 19 MS | | | Cascaded 3D CNNs | | | |
| Roca [127] | Predict EDSS Score of MS Patients | OFSEP Cohort | MRI | DS1: 480 | -- | DA | 3D-CNN | TensorFlow | -- | MSE=3 |
| | | | | DS2: 491 | | | | | | |
| | | | | DS3: 475 | | | | | | |
| Nair [129] | Uncertainty Estimates | Clinical | MRI | 1064 MS | -- | -- | 3D-CNN | -- | -- | -- |
| Brown [132] | Segmentation and Calibration | CPDDS | MRI | 256 Participants | -- | -- | 3D-CNN | Theano | -- | Mean J=74 |
| Sepahvand [152] | Future Disease Activity Prediction | Clinical | MRI | 1068 MS | -- | -- | 3D CNN | -- | -- | Acc=80.21<br>Sen=80.11<br>Spec=79.16<br>Prec=91.82 |
| | Segmentation | | | | | | Modified U-Net | | | |
| Rosa [153] | Segmentation | Clinical | MRI | 105 MS | FSL | Manual Segmentation, LeMan-PV | Cascade of Two 3D Patch-Wise CNNs | -- | -- | DSC=60<br>VD=40 |
| Tousignant [162] | Prediction of Disability Progression of MS Patients | Clinical | MRI | 465 MS | -- | 2 Lesion Masks | 3D-CNN | -- | 4 | AUC=70.1 |
| Yoo [163] | Predicting Future Disease Activity in Patients with Early Symptoms Of MS | Clinical | MRI | 140 Subjects | -- | DA | 3D-CNN | Theano, cuDNN | 7 | Acc=72.9<br>Sen=78.6<br>Spec=65.1<br>AUC=71.8 |
| Kazancli [168] | Lesion Segmentation and Classification | Clinical | MRI | 59 MS | FreeSurfer | Patch Extraction | Two 3D-CNNs in a Cascade Fashion | TensorFlow | -- | DSC=57.5 |
| Gros [169] | Segmentation of The Spinal Cord and Lesions | Clinical | MRI | 459 HC, 471 MS, 112 With Other Spinal Pathologies | FSL | Manual Segmentation | Sequence of Two CNNs | Keras, TensorFlow | -- | MSE=1 |
| | | | | | | | | | | DSC=94.6 |
| | | | | | | | | | | DSC=60.4 |
| Valverde [173] | Lesion Segmentation | MICCAI 2008 | MRI | 60 Patients | FSL | -- | 3D-CNN | Keras, TensorFlow | -- | DSC= 63<br>Sen= 55<br>Pre= 79<br>Score= 91.33 |
| | | MICCAI 2016 | | | | | | | | |
| | | ISBI 2015 | | | | | | | | |
| | | Clinical | | | | | | | | |
| Zhang [176] | MS Identification | eHealth Laboratory and Private Data | MRI | 38 MS, 26 HC | -- | HS, DA | 3D-CNN | -- | -- | Acc=98.23<br>Sen=98.22<br>Spec=98.24 |
| Yoo [178] | Predicting Conversion to MS from CIS | Clinical | MRI | 140 Subjects | -- | DA | 3D-CNN | Theano | 7 | Acc=75<br>Sen=78.7<br>Spec=70.4 |
| Eitel [188] | Classification | Clinical | MRI | 76 MS, 71 HC | FSL | DA | 3D-CNN | Keras, TensorFlow | -- | Acc=87.04<br>AUC=96.08 |
| Valverde [192] | Lesion Detection and Segmentation | MICCAI 2016 | MRI | 53 MS | -- | 3D Patch Extraction | 3D-CNN | Theano | -- | -- |
| Valverde [175] | White Matter Lesion Segmentation | MICCAI 2008 | MRI | 45 MS | FSL, SPM | Patch Extraction, DA | Cascade of Two 3D Patch-Wise CNNs | Theano | -- | VD=40.8<br>TPR=68.7<br>FPR=46 |
| | | Clinical | | 60 MS | | | | | | |
| Gessert [170] | Lesion Segmentation | Clinical | MRI | 89 MS | -- | Lesions Extraction | Attention-Guided Two-Path CNNs | -- | 3 | LFPR=26.4<br>LTPR=74.2<br>DSC=62.2 |
| | | | | 33 MS | | | | | | |
| Sepahvand [116] | Segmentation and Detection | Clinical | MRI | 886 MS | -- | -- | NE SubNet | -- | 5 | Sen=97.74<br>Spec=69.26<br>AUC=90.83 |
| McKinley [121] | Lesion Quantification | Bern | MRI | 26 MS | FreeSurfer | -- | DeepSCAN | -- | -- | Acc=85<br>AUC =99.9 |
| | | Zurich | | 8 MS | | | | | | |
| | | Munich | | -- | | | | | | |
| Ackaouy [122] | Segmentation | MICCAI 2016 | MRI | 53 Images of MS Patients | -- | -- | Seg-JDOT | Keras | -- | -- |
| Maggi [128] | CVS Assessment in White Matter MS Lesions | Multicenter Cohort | MRI | 42 MS, 33 MS Mimics, 5 Uncertain Diagnosis | -- | DA | CVSnet | Keras, TensorFlow | 10 | Lesion-Wise Median Balanced Acc=81<br>Subject-Wise Balanced Acc= 89 |
| McKinley [227] | Simultaneous Lesion and Brain Segmentation | MSSEG 2016 | MRI | 15 Datasets | FSL | -- | DeepSCAN | -- | -- | DSC=60<br>F1-S=57 |
| | | Insel90 | | 90 Datasets | | | | | | |
| | | Insel32 | | 32 Patients | | | | | | |
| HASHEMI [189] | Data Imbalance | MICCAI 2016 | MRI | 53 MS | -- | Patch Extraction | 3D Patch-Wise FC-Dense-Net | -- | 5 | DSC=69.9 |
| | | ISBI 2015 | | 19 MS | | | | | | DSC=65.74 |
| McKinley [171] | Lesion Segmentation | Insel90 | MRI | 90 Datasets | Freesurfer, FSL | Manual Segmentation Weak Label | DeepSCAN | -- | -- | DSC=59<br>Sen=50<br>Pre=68 |
| | | Insel32 | | 32 MS | | | | | | |
| Vincent [114] | Segmentation | Clinical | MRI | 642 MS | -- | -- | FiLMed-Unet | PyTorch | -- | DSC=72 |
| Vang [130] | Segmentation | Clinical | MRI | 261 Patients | LST | -- | Synergy-Net | -- | -- | DSC=61.52<br>Prec=42.27<br>Sen=59.11 |
| | | ISBI 2015 | | 5 Patients | | | | | | |
| Calimeri [193] | Classification of MS into 4 Clinical Profiles CIS, RR, SP, PP | Clinical | MRI | 90 MS | -- | Brain Structural Connectivity Graph | Graph Based Neural Networks | -- | 10 | Prec=82<br>Recall=79<br>F1-S=80 |
| Marzullo [186] | Classification into 4 Clinical Profiles | Clinical | MRI, DTI | 90 MS (12 CIS, 30 RRMS, 28 SPMS, 20 PPMS), 24 HC | -- | Brain Structural Connectivity Graph, Graph Local Features | Graph Convolutional Neural Network (GCNN) | -- | 3 | Pre=92<br>Recall=92<br>F1-S=92 |
| Dai [187] | Compressed Sensing MRI | eHealth Laboratory | MRI | 500 Images | MATLAB | 3 Sampling Masks | MDN | Caffe, PyTorch, TensorFlow | -- | PSNR= 38.73<br>SSIM=98.6 |
| Dewey [183] | Contrast Harmonization Across Scanner Changes | Clinical | MRI | 10 MS<br>2 HC<br>45 MS | -- | Super-Resolved and Anti-Aliased, Gain-Correction | DeepHarmony | Keras, TensorFlow | 6 | -- |
| Yoo [161] | Distinguishing NMOSD from MS | Clinical | MRI, DWI | 82 NMOSD, 52 MS | -- | -- | Hierarchical Multimodal Fusion (HMF) Model | -- | 7 | Acc= 81.3<br>Sen= 85.3<br>Spec= 75<br>AUC= 80.1 |
| Essa [134] | Segmentation | MICCAI 2008 | MRI | 45 Scans | -- | 3D Patches Extraction | 2 Parallel R-CNN | -- | -- | Sen=61.8 |
| Hou [148] | Segmentation | ISBI 2015 | MRI | 19 MS | -- | DA | Cross Attention Densely-Connected Network (CA-DCN) | Keras, TensorFlow | -- | DSC=64.3<br>LFPR=10.5<br>LTPR=441 |
| Ulloa [150] | Segmentation | ISBI 2015 | MRI | 19 MS | -- | Circular Non-Uniform Sampling Patch, DA | Single-View Multi-Channel (SVMC) | Keras, TensorFlow | 5 | DSC=67.10 |
| Zhang [151] | Segmentation | Clinical | MRI | 43 MS | FSL | -- | Recurrent Slice-Wise Attention Network (RSANet) | PyTorch | -- | Sample avg. dice= 66.011<br>Voxel avg. dice= 71.054<br>Sample avg. IoU= 50.917<br>Voxel avg. IoU= 55.201 |
| Narayana [117] | Classification | Clinical | MRI | 1008 MS | -- | DA | VGG16+FCN | Keras, TensorFlow | 5 | Acc=70<br>Sen=72<br>Spec=70 |
| Barquero [133] | Classification of rim+/rim- Lesions | Clinical | MRI | 124 MS | FreeSurfer, FSL | Different Methods, DA | RimNet (two parallel CNNs inspired by VGGNet) | -- | 4 | Acc= 93.8<br>Sen= 75.8<br>Spec= 95.1<br>F1-S= 62.3 |
| Ye [140] | Classification | Clinical | MRI | 38 MS | -- | DTI and DBSI Analyses | DNN | MATLAB, TensorFlow | -- | Acc= 93.4<br>Sen= 99.1 |
| | | | | | | | | | | Spec= 97.3<br>F1-S= 97.3<br>AUC=99.8 |
| Fenneteau [167] | Lesions Segmentation | Different Datasets | MRI | Different Cases | FSL | Patch Extraction | 3D U-net | Keras, TensorFlow | -- | DSC=67.63<br>Sen=61.47<br>Pre=79.30 |
| Coronado [119] | Segmentation and Detection | CombiRx | MRI | 1006 MS | -- | MRIAP Pipeline | 3D U-Net | -- | -- | DSC=77 |
| Narayana [120] | Segmentation | CombiRx | MRI | 1008 MS | -- | MRIAP Pipeline | Multiclass U-Net | Keras, TensorFlow | -- | GM-DSC=94<br>WM-DSC=94<br>CSF-DSC=96<br>Lesion-DSC= 86 |
| Rosa [123] | Segmentation and Detection | Clinical | MRI | 60 MS | -- | DA | Multi-task 3D U-Net + ICD | TensorFlow | 6 | Acc=86 |
| Rosa [124] | Segmentation | Basel University Hospital | MRI | 54 MS | NiftyNet | DA | 3D U-Net | TensorFlow | 6 | Detection Rate=76<br>DSC=60 |
| | | Lausanne University Hospital | | 36 MS | | | | | | |
| Narayana [126] | Segmentation | CombiRx | MRI | 1008 MS | -- | MRIAP Pipeline, DA | 2D U-Net | Keras, TensorFlow | -- | DSC= 90<br>TPR= 81<br>FPR= 28 |
| Abolvardi [141] | Registration Based Data Augmentation | Longitudinal MS lesion Dataset | MRI | 19 MS | -- | -- | 3D U-Net | -- | 5 | DSC=61.4 |
| Falvo [142] | Accelerating MRI | Public Dataset | MRI | 30 MS | -- | -- | Multimodal Dense U-Net (MDU) | MATLAB, Keras | -- | Acc=97 |
| Ghosal [143] | Segmentation | MICCAI 2016 | MRI | 15 MS | -- | -- | Light Weighted U-Net | Keras, TensorFlow | 5 | Acc=96.79<br>Sen=65<br>Spec=86<br>DSC=76 |
| Kumar [144] | Segmentation | MICCAI 2016 | MRI | 15 MS | -- | DA | Modified Dense U-Net | Keras | 5 | DSC=86.6<br>Sen=85.6 |
| Kats [146] | Segmentation | ISBI 2015 | MRI | 19 MS | -- | Soft Labeled Mask | 2D U-Net Based FCNN | -- | 5 | DSC=57.8<br>Prec=83.8<br>Recall=46.6 |
| Feng [147] | Segmentation | ISBI 2015 | MRI | 19 MS | -- | Different Methods | 3D U-Net | -- | -- | DSC=68.4 |
| Narayana [157] | Tissue Classification | CombiRx | MRI | 1008 MS | -- | MRIAP Pipeline, DA | Multi-Class U-Net | Keras, TensorFlow | -- | WM-DSC=94<br>GM-DSC=94<br>CSF-DSC=97<br>Lesion-DSC= 85 |
| Hu [160] | Lesion Segmentation | ISBI 2015 | MRI | 19 MS | -- | Data Enhancement | 3D Attention Context U-Net (ACU-Net) | Keras, TensorFlow | -- | DSC= 63.45<br>PPV= 86.82 |
| | | | | | | | | | | LTPR= 47.87<br>LFPR= 12.99 |
| Gabr [164] | Brain and Lesion Segmentation | CombiRx | MRI | 1008 MS | -- | MRIAP Pipeline, DA | Multiclass U-Net FCNN | Keras, TensorFlow | -- | WM-DSC= 95<br>GM-DSC=96<br>CSF-DSC=99<br>T2 Lesions-DSC=82 |
| Salem [137] | Segmentation | VH dataset | MRI | 60 MS | ROBEX, ITK, Nifty Reg | 3D Patch Extraction | FCNN | Keras, TensorFlow | -- | DSCs=55<br>DSCd=83 |
| Yoo [177] | Distinguish Between MS Patients and HC | Clinical | MRI | 55 MS<br>44 HC | FSL | Lesion Masks, Patch Extraction | Multimodal Deep Learning Network | -- | 11 | Acc= 87.9<br>Sen=87.3<br>Spec=88.6 |
| Sujit [145] | Automatically Evaluate the Quality of Multicenter Structural Brain MRI Images | ABIDE<br>CombiRx | MRI | 1112 Subjects<br>110 MS | SPM | DA | Ensemble DL Model | Keras, TensorFlow | -- | Acc=84<br>Sen=77<br>Spec=85<br>AUC=90 |
| Finck [115] | Produce synthDIR | Clinical | MRI | 100 MS | -- | -- | DiamondGAN | -- | -- | Detection Rate=31.4<br>CNR=22 |
| Wei [125] | Predicting PET-Derived Myelin Content from Multi Sequence MRI | Clinical | MRI, PET | 18 MS, 10 HC | FSL, FreeSurfer | DA, Lesion-Filling Procedure | Conditional Flexible Self-Attention GAN (two CF-SAGAN used as Sketcher and Refiner) | TensorFlow | 3 | DSC= 91 |
| Shaul [138] | Subsampled Brain MRI Reconstruction | ISBI 2015 | MRI | 80 MS | -- | Inverse Orthonormal 2D FT | GAN | -- | -- | PSNR=28.26<br>SSIM=90<br>DSC=90.4 |
| Zhang [180] | Lesion Segmentation | Clinical | MRI | 69 MS | FSL | DA, Pseudo 3D Slice Extraction | MS-GAN | PyTorch | -- | DSC=67.2<br>Recall=69.2<br>Prec=72.4 |
| Wei [182] | Predicting PET-Derived Demyelination from Multimodal MRI | Clinical | MRI, PET | 18 MS, 10 HC | FSL | ROIs Extraction | Sketcher-Refiner GANs | Keras, Theano | 3 | MSE=0.0083<br>PSNR=30.044 |
| Wei [185] | Predict The PET-Derived Myelin Content Map from a Combination of MRI Modalities | Clinical | MRI, PET | 18 MS, 10 HC | -- | ROIs Extraction | Sketcher-Refiner GANs | Keras | 3 | -- |
| Hagiwara [224] | Improving the Quality of Synthetic FLAIR Images | Clinical | MRI | 40 MS | SyMRI Software, FSL | -- | Conditional GAN | Python, Chainer | -- | PSNR= 35.9 NRMSE=27 |
| Karaca [166] | Classification of MS Subgroups | Clinical | MRI | 120 MS | | Lesion Diameter Data | SSAE | MATLAB | 10 | Acc=99.78 |
| Vogelsanger [226] | Latent Space Analysis | Clinical | MRI | 616 MS | ITK, Framework, FSL | Trimming and Down Sampling, Bounding the Voxel Values | Introspective Variational Autoencoder (intro-VAE or IVAE) | Keras, TensorFlow | -- | Pre=92 Recall= 89 |
| Aslani [118] | Segmentation | Clinical | MRI | 117 MS | FSL | -- | Traditional Encoder-Decoder Network with Regularization Network | Keras, TensorFlow | 5 | DSC=50 |
| McKinley [191] | Lesion Segmentation | MICCAI 2016 | MRI | 53 MS | -- | Lesion Mask | Nabla-Net | Keras, Theano | -- | -- |
| Krüger [136] | Segmentation | Different Dataset | MRI | Different Number of Cases | SPM, LST | DA | Fully 3D Convolutional Encoder-Decoder Architecture | -- | -- | Sen=60 DSC=45 |
| Brosch [174] | Lesion Segmentation | MICCAI 2008<br>ISBI 2015<br>clinical | MRI | 43 MS<br>21 MS<br>195 MS | FSL | Ground Truth Segmentations Via Semiautomatic 2D Region-Growing Technique | Convolutional Encoder Network with Shortcut Connections (CEN-s) | -- | -- | DSC= 63.83 LTPR= 62.49 LFPR= 36.10 VD= 32.89 |
| Tripathi [139] | Denoising Of MRI Scans | University of Syprus Dataset | MRI | -- | -- | -- | CNN-DMRI | TensorFlow | -- | PSNR= 38.51 SSIM=97 |
| Gessert [113] | Segmentation | University Hospital of Zurich, Switzerland | MRI | 44 MS | -- | -- | Enc-convGRU-Dec | -- | -- | DSC=64 LTPR=84 |
| Andermatt [131] | Segmentation | ISBI 2015 | MRI | 20 MS | -- | DA | MD-GRU | -- | -- | DSC= 62.85 HD= 32.60 AVD= 1.83 |
| Sander [194] | Brainstem Segmentation | Clinical | MRI | Different Number of Cases | -- | DA | MD-GRU | -- | -- | DSC=98 |

## 3. Discussion

The purpose of this paper is to provide a complete overview of works done in the field of automated MS diagnosis using MRI modalities and DL techniques. In this review paper, comprehensive details of most of the works carried out are provided for readers. Table (2) summarizes works done in the field of MS detection using DL techniques and MRI modalities. According to Table (2), the discussion section is organized into several subsections. Subsections of the discussion comprises of comparing conventional machine learning techniques with DL, types of MS diagnosis applications, datasets, MRI modalities in MS diagnosis, MRI preprocessing toolboxes, DL architectures, DL toolboxes, and finally classification algorithms in the diagnosis of MS. Also, the details of DL networks developed for MS diagnosis are provided in the appendix table A. They are briefly presented in the following sections.

### 3.1. Comparison of deep learning and conventional machine learning methods

Research in the field of MS diagnosis using AI techniques is divided into two categories: conventional machine learning and DL. In references [27-29], various conventional machine learning works developed to diagnose MS from modalities are provided. In CADS based on conventional machine learning, the main aim is to combine different algorithms together (including preprocessing to classification) to achieve highest accuracy. This is a relatively complex task and requires a great deal of knowledge in the field of machine learning [93].

In recent years, DL techniques have received a particular position in MS diagnosis [238]. Unlike conventional machine learning approaches, DL techniques are highly effective in diagnosing MS. In CADS based on DL methods, deep layers are exploited to extract the features [238]. This enhanced the effectiveness of CADS in MS diagnosis. The application of DL techniques have yielded hope for accurate diagnosis of MS using MRI modalities.

### 3.2. Comparison of deep learning applications for diagnosis of MS

In this section, a comparison between various applications based on DL methods for diagnosing MS are presented. It can be noted from Table (2) that the majority of works have been focused on segmentation and classification approaches, or combination of both as shown in Figure (10). It can be noted from the Figure that it is important to perform segmentation to diagnose MS. This will help to identify the MS lesions on MR images. The DL techniques can be used to recognize the exact location and dimensions of the MS lesion, which will help the clinicians to confirm their diagnosis. There are datasets available with manual segmentation of MR images. This has paved the way to have more segmentation works being done using DL techniques with MRI neuroimaging modalities.

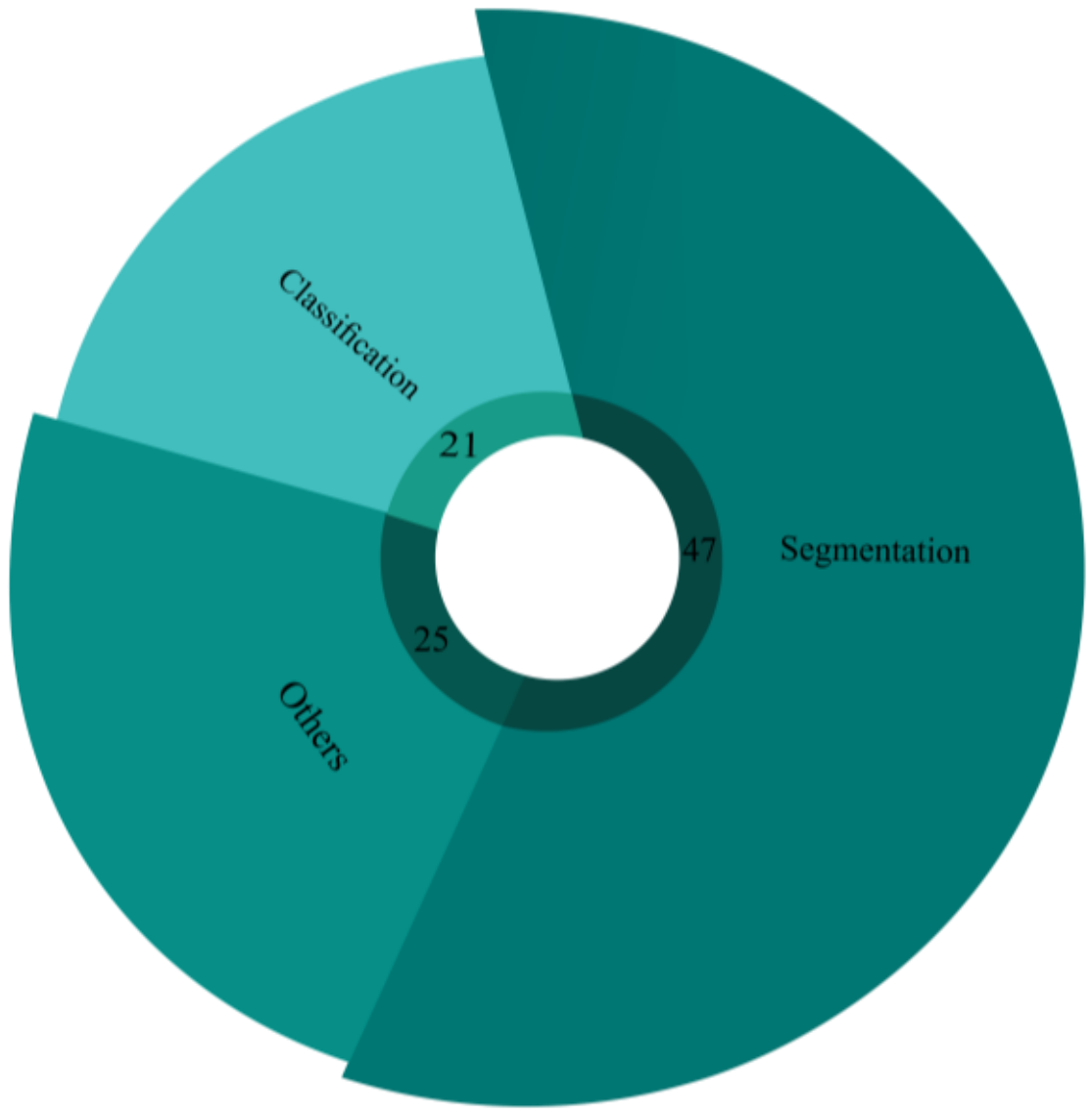

Fig. 10. Number of Applications used for MS diagnosis.

### 3.3. Comparison of available MRI datasets for diagnosis of MS

Several available datasets of MRI modalities have been introduced for the automated diagnosis of MS. The available MRI datasets for the automated MS diagnosis are given in Table (1). In Table (2), the clinical datasets on diagnosis of MS are listed. Figure (11) illustrates the number of datasets employed to diagnose MS. It can be seen, that most of the works have used clinical data. Among the available datasets, ISBI 2015 is the most frequently used one for MS diagnosis research. This dataset contains different types of sMRI modalities, and  motivated its application in majority of relevant studies.

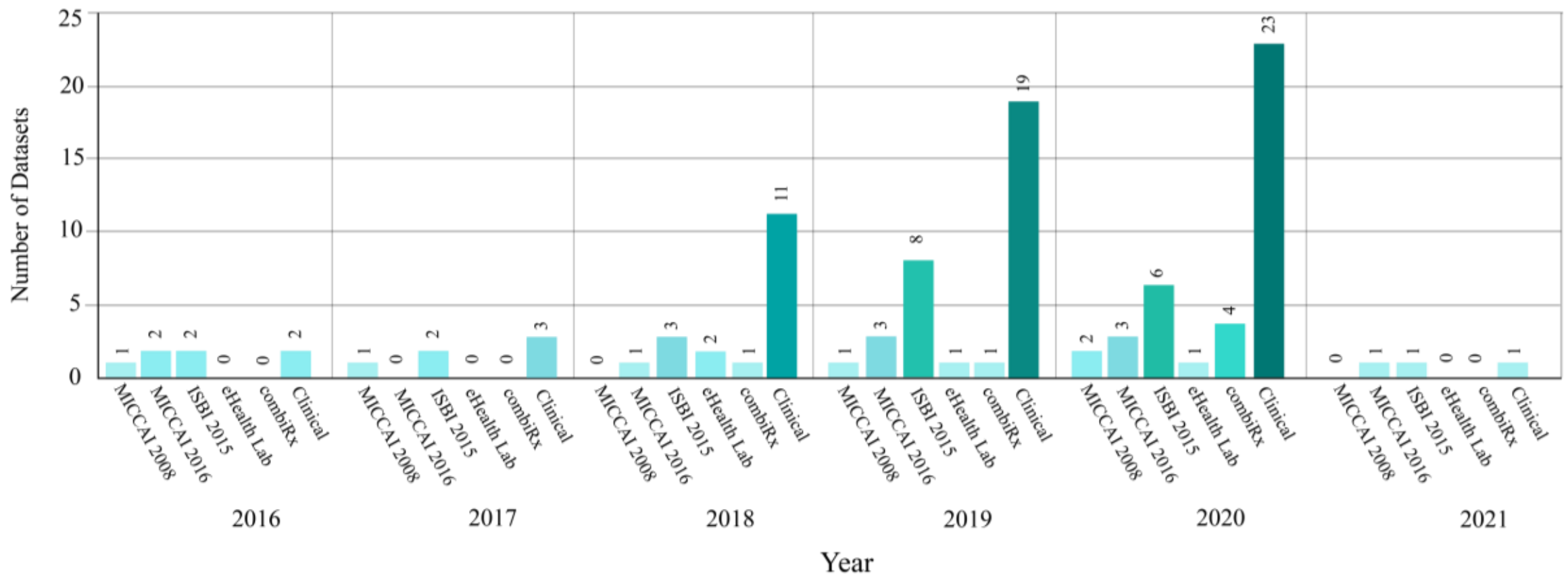

Fig. 11. Number of datasets used for MS diagnosis.

### 3.4. Comparison of MRI neuroimaging modalities for diagnosis of MS

Previous sections explained different types of neuroimaging modalities for MS diagnosis. As presented in Table (1), so far, datasets with sMRI modalities have been provided for research purposes. MRI neuroimaging modalities for the diagnosis of MS are explained in another section of Table (2). It can be noted from Table (2) that the number of MRI modalities used in the MS diagnosis is depicted in Figure (12). It can be seen from this Figure, that the use of sMRI modalities to diagnose MS has grown more than other neuroimaging modalities. Based on Figure 12, few researchers have employed PET imaging in addition to sMRI modalities for MS diagnosis, thereby enhancing the precision and efficiency of CADS.

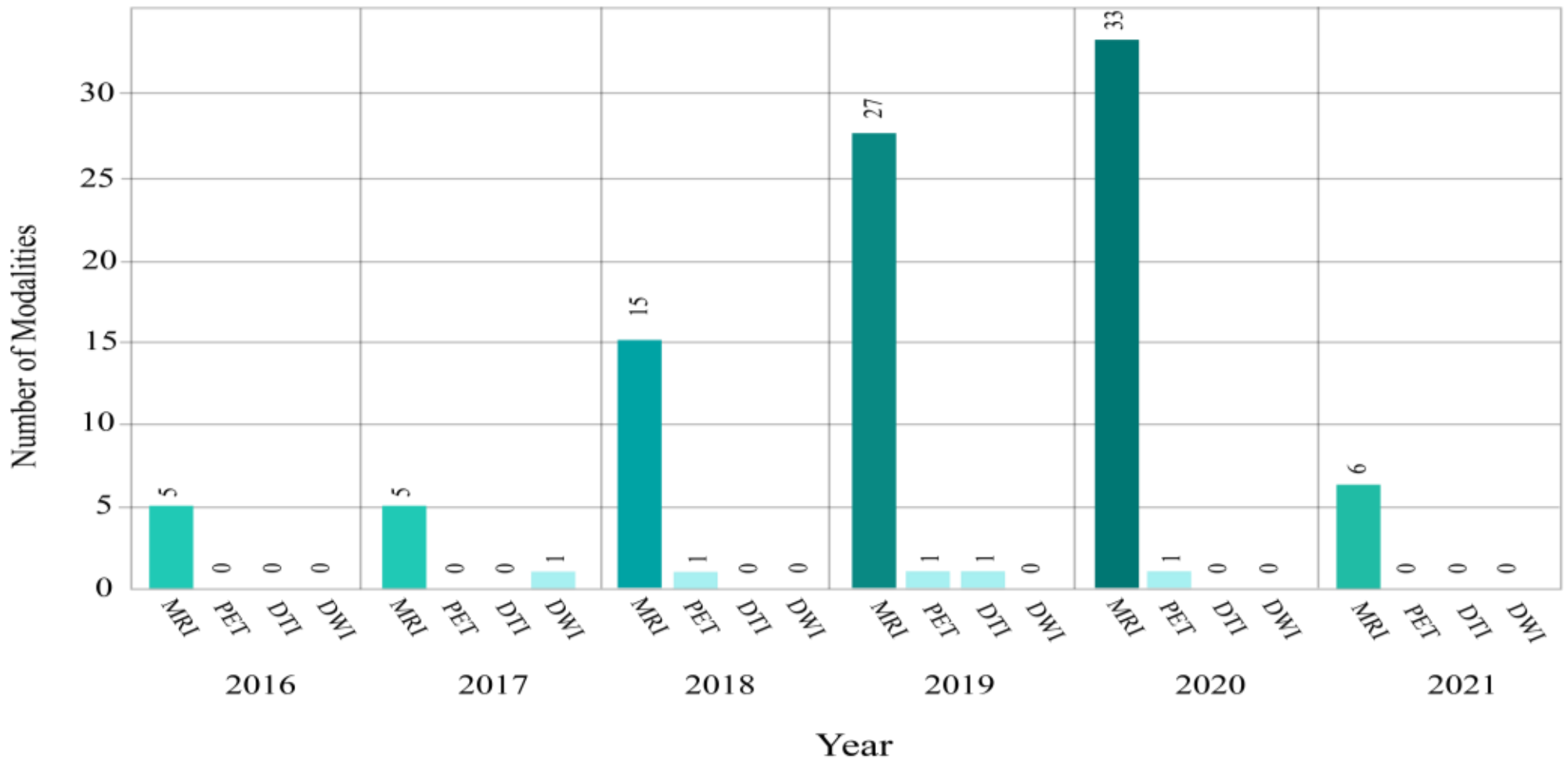

Fig. 12. Number of neuroimaging modalities used in the MS diagnosis.

### 3.4. Comparison of various MRI preprocessing toolboxes for diagnosis of MS

As mentioned in the previous sections, preprocessing is an important step in MRI. The pre-processing of MRI modalities has certain stages, as presented in Section 2.2. The implementation of these pre-processing steps are usually time-consuming, and several toolboxes have been proposed to overcome this problem. Various toolboxes have been used for low-level preprocessing of MRI modalities, the most important of which are the FMRIB software library (FSL) [195], FreeSurfer [196], statistical parametric mapping (SPM) [197], and Matlab. The number of MRI preprocessing toolboxes used to detect MS are shown in Figure (13). It can be seen, that the FSL toolbox is widely used in many works.

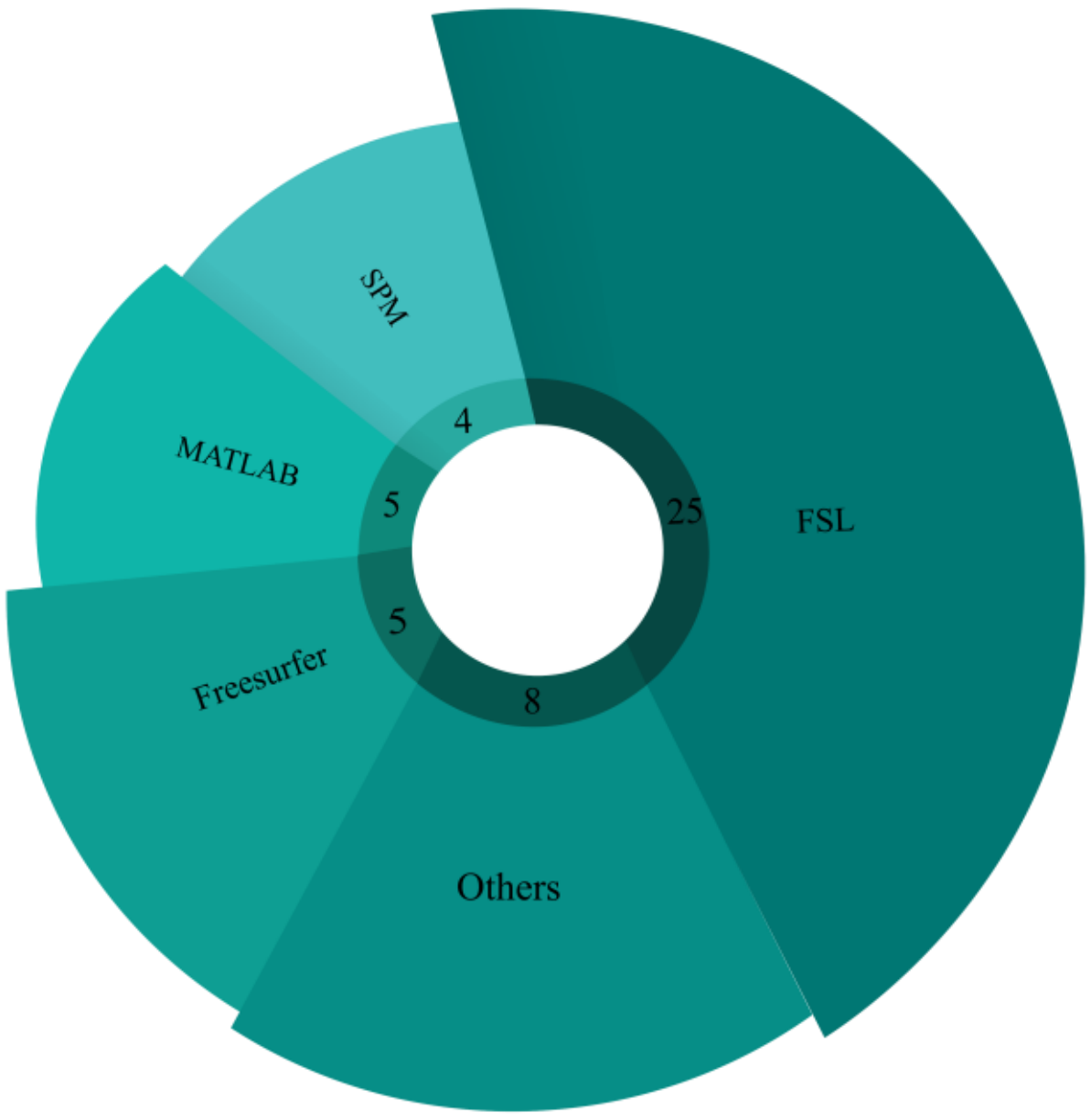

Fig. 13. Number of MR images preprocessing toolboxes used for MS diagnosis.

### 3.5. Comparison of different DL methods for diagnosis of MS

This review mainly examines different DL methods developed for MS diagnosis. The most well-known DL techniques for MS diagnosis based on MRI modalities are presented in Section 2.3. Based on Section 2.3, DL segmentation and classification methods have been employed for MS diagnosis. Among DL methods, only CNN models are employed in different types of segmentation and classification techniques. The type of DL model used for automated MS detection using MRI modalities are given in Table (2). The number of DL networks used every year for MS diagnosis is shown in Figure (14). It can be noted from this Figure that CNN models have been widely used to diagnose MS from MRI modalities. The popularity of CNN models compared to other DL techniques lies in getting high performance using brain MR images.

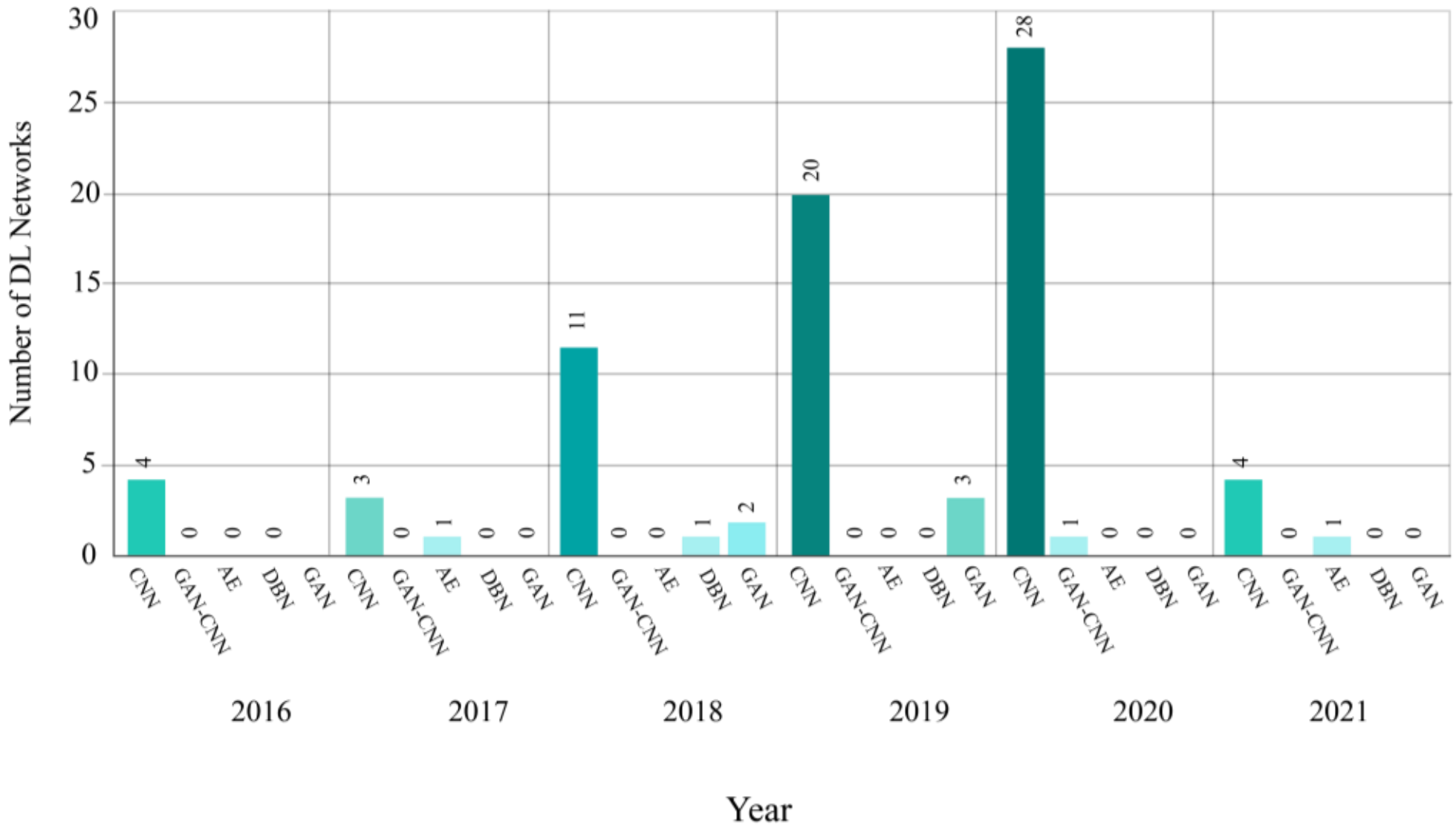

Fig. 14. Types of DL networks used for MS diagnosis.

### 3.6. Comparison of different DL toolboxes for diagnosis of MS

Numerous toolboxes have been provided for implementing DL models by companies such as Google or Facebook. The various tools used to develop DL architectures are shown in Table (2). The most important DL tools are TensorFlow, Keras, Caffe, and PyTorch [198-200]. Various DL toolboxes used by authors are also shown in Table (2). The number of DL toolboxes used in automated MS diagnosis is displayed in Figure (15). It can be noted from Figure (15) that the Keras toolbox is the most used system to MS detection using MRI modalities. Keras is a powerful, free-of-charge, easy-to-use, and open-source library for the development and evaluation of DL models. It covers one numerical machine learning library of TensorFlow, and allows researchers to easily implement DL models. Hence, Keras is the most popular library among DL researchers for diagnosis of MS.

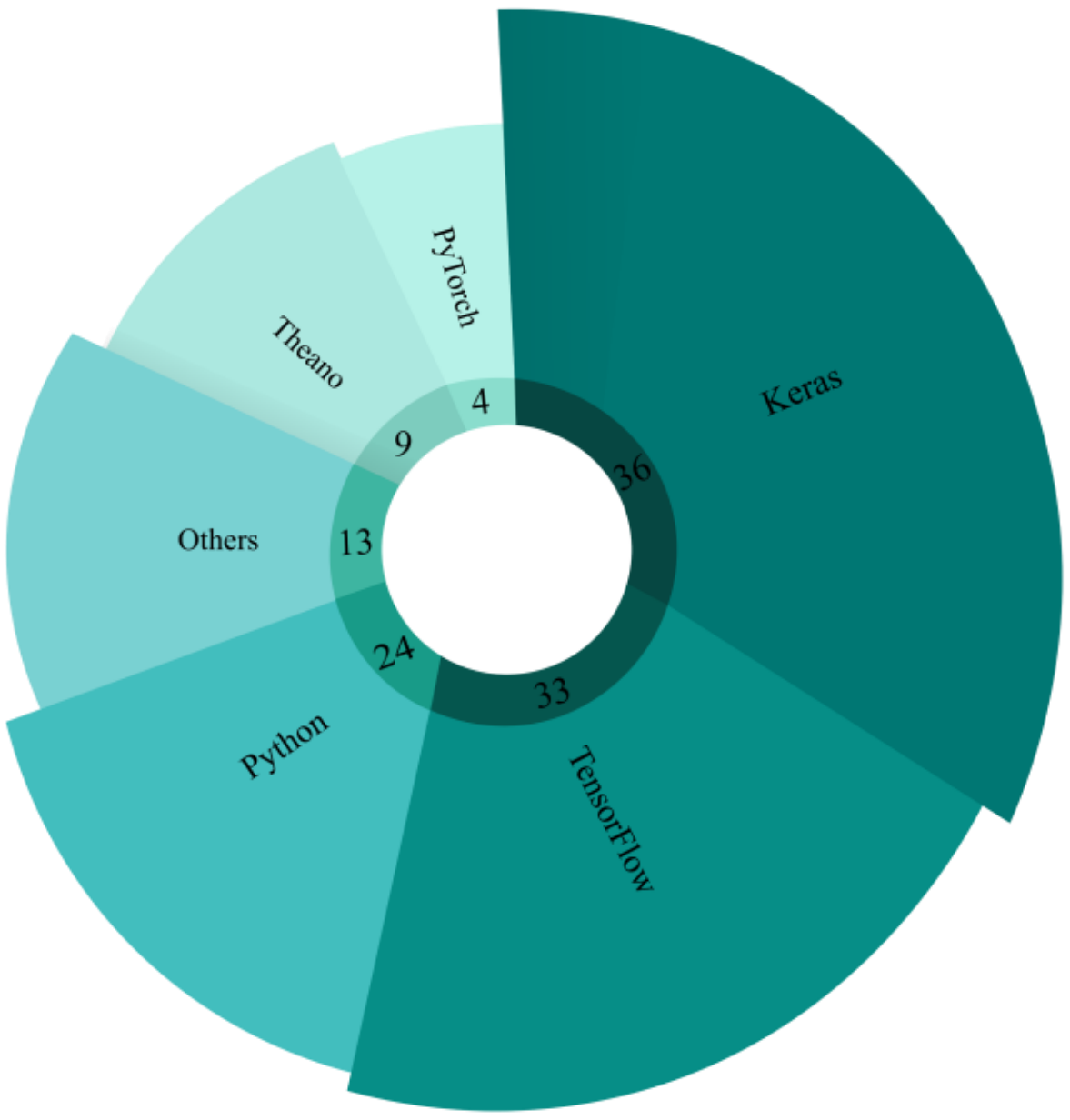

Fig. 15. Types of DL toolboxes used for MS diagnosis.

### 3.7. Comparison of classification methods for diagnosis of MS

The activation function of the last layer used for classification in DL models is the last part of the DL-based CADS shown in Table (2). The number of activation functions used in DL-based CADS for MS detection is shown in Figure (16). It can be noted that, the softmax function has yielded the highest classification performance.

## 4. Challenges

In this section, the most important challenges in the automated MS diagnosis using MRI neuroimaging modalities and DL techniques are discussed. The inaccessibility of available sMRI databases belonging to more subjects and different modalities is the first challenge. The second challenge is the inaccessibility of datasets with functional neuroimaging modalities for MS diagnosis research. Finally, DL models and hardware resources remain the third challenge. These challenges are discussed below.

### 4.1. Unavailable big data sMRI datasets with different modalities

In the automated MS diagnosis, huge datasets are needed to obtain highest classification performance. The datasets available have a finite number of subjects and therefore advanced DL models cannot be employed to investigate them. In segmentation applications, the principal objective is to apply a DL method to delineate the MS lesions in the MRI modalities. This can be achieved when the DL network is first trained using a huge number of MR images. This huge number of images obtained from large number of subjects need to be manually delineated and fed as input to perform automated segmentation.

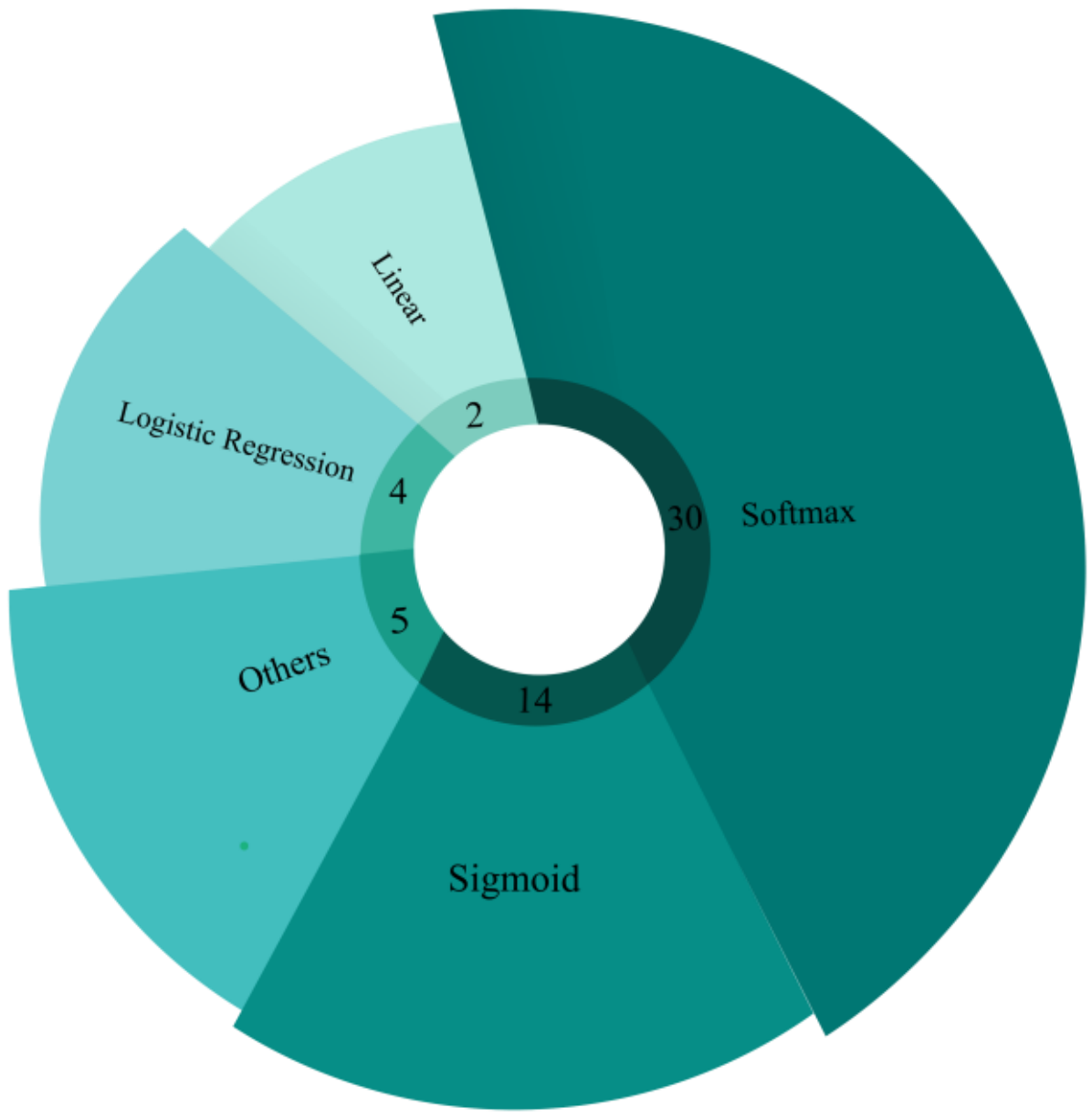

Fig. 16. Number of MS classification works proposed using DL methods.

**4.2. Unavailable functional Neuroimaging modalities for MS diagnosis**

This section addresses the most important challenges of functional neuroimaging modalities in the diagnosis of MS. The most important challenges include the inaccessibility of available fMRI datasets and other functional neuroimaging datasets. They are briefly discussed below.

**(1). Unavailable fMRI datasets**

The unavailability of functional MRI (fMRI) datasets is the key challenge. fMRI modalities yield important information about brain function to specialist physicians and neurologists [60]. fMRI modalities contain two categories: rs-fMRI and Task-fMRI [60]. Specialists have concluded that fMRI modalities are effective in diagnosing brain diseases, including MS [239]. In clinical studies, there have been much debate on the importance of using fMRI to automatically diagnose MS [201-203]. Hluštík [239] in a study indicated that applying fMRI modalities in the diagnosis of MS is of great significance. In this study, motor, visual and cognitive networks in MS patients were examined using fMRI modalities. One of the advantages of fMRI modalities is that they can determine the location of MS based on the functioning of brain neurons. But fMRI modalities are more complex than sMRI modalities. This has led to research into the diagnosis of MS using fMRI modalities and AI techniques. In [204-205], A few researchers have taken the advantage of fMRI neuroimaging modalities alongside AI techniques to detect MS and have achieved satisfactory results. Unfortunately, the lack of access to fMRI datasets involving large number of subjects has prevented researchers from developing accurate and robust DL techniques to diagnose MS. The availability of fMRI datasets with large number of subjects helps the researchers to develop an accurate MS diagnosis model which can assist the physicians to confirm their manual screening.

**(2). Unavailable other functional neuroimaging datasets**

Furthermore, the lack of accessibility to datasets from other structural neuroimaging modalities is another challenge. Few clinical studies have used electroencephalogram (EEG), functional near-infrared spectroscopy (fNRIS), and magnetoencephalography (MEG) to diagnose MS [206-210]. Also, few authors have used EEG signals with conventional machine learning algorithms to detect MS [211-213] automatically.

Recently many multimodality techniques have been proposed to accurately diagnose brain disorders with satisfactory outcomes [214-216]. Very few clinical works have been conducted to diagnose MS using multimodality techniques such as EEG-fMRI [217] and MEG-fMRI [218]. The performance of the system can be improved by using data fusion techniques.

### 4.3. Dl methods and hardware's

Another challenge is the selection of DL approach and hardware resources. The development of a DL method to distinguish MS using MRI modalities demands more images and experience. Lack of access to adequate hardware resources to implement DL architectures is another big challenge. Although servers such as Google Colab, Amazon, and cater good hardware resources are available for researchers to train DL networks, using these servers in the real world is a big challenge and rises privacy concerns.

## 5. Future directions in the automated MS diagnosis using DL techniques

In this section, directions for future work on MS diagnosis based on MRI neuroimaging modalities and DL techniques are delineated. Future research directions include three categories: datasets, application of novel DL models, and rehabilitation systems for MS patients.

### 5.1. Future works in datasets for automatic diagnosis of MS

The available datasets for MS diagnosis are presented in Table (1). It can be noted from Table (1) that most of the available sMRI datasets are small (limited subjects). Hence, the developed automated systems for MS diagnosis may not be accurate and robust. As a future research direction, proposing sMRI datasets with large number of subjects will help to design a practical software for MS diagnosis. Lack of access to datasets of functional modalities such as fMRI is another challenge in MS diagnosis. As a future work, we propose to have more accessible datasets of fMRI modalities for conducting research on MS diagnosis. This can pave the way for developing more accurate systems for screening of brain function of MS patients by using DL techniques and fMRI modalities.

In [125], the PET functional modality is used for MS diagnosis. In the future, presenting accessible datasets of PET modality can contribute to conducting applied studies in this domain. Physicians in clinical research have utilized combined modalities such as PET and MRI to diagnose brain diseases [240-242]. Free available of such combined datasets (PET and MRI modalities) can be used to develop automated systems for MS diagnosis.

MRI modalities have been used to classify or segment MS lesions [223]. The T1-w modality, however, is often used to segment the brain tissue. MS lesions are usually manifest as hyperintensities in the T2-w and PD-w modalities [223]. The major drawback of these two modalities are the similarity in lesion intensities and CSF, which makes the segmentation difficult. In such cases, the T2-FLAIR modality can be of great significance, but this modality becomes problematic when dealing with subcortical structures [223]. Therefore, in future, DL models with several neuroimaging modalities can help to identify and segment the lesion of MS disease.

### 5.2. Future works in DL methods for automatic diagnosis of MS

Recently, GAN models have been introduced in medical applications and a lot of research is being done in this field [228-230]. As mentioned, the lack of medical data is an obstacle for training DL networks. GAN models have mostly been able to address the lack of medical data to train DL networks [228-230]. For this purpose, various GANs can be employed to generate large amounts of MRI modalities in the future works of MS diagnosis. Additionally, some novel models such as graph theory-based architectures [231, 232], zero-shot learning [233-235], and representation learning [236-237] can be used by researchers as future works in MS diagnosis using MRI neuroimaging modalities.

### 5.3. Future works in design of rehabilitation systems for automatic diagnosis of MS

Cloud computing is a novel medical technology that has attracted considerable attention from researchers [243-245]. It allows researchers to store large MRI data in a cloud space. DL methods can also be implemented and simulated in the cloud space. It is expected that future works employ cloud computing to study MS diagnosis based on MRI modalities and DL methods.

The Internet of Things (IoT) is another developing technology in the medical industry [246-248]. Access to neurologists is challenging in most treatment centers. Thus, in the future, the use of IoT and DL technologies can facilitate the process of treatment and diagnosis for MS patients.

## 6. Conclusion

MS is a chronic disease that directly attacks the central nervous system, including the brain, spinal cord, and optic nerves. Early diagnosis of MS is of great significance as it can prevent the progression of the disease and save life. MRI neuroimaging modalities provide important information about brain tissue and structure to specialist physicians. Therefore, MRI modalities are widely used to obtain the presence of MS lesions. Various methods have been proposed to diagnose the MS using MRI modalities and machine learning techniques. In this paper, different components of CADS employed for MS diagnosis using DL, and automated MS detection systems developed are presented in Table (2).

The works done on MS diagnosis using MRI modalities and DL techniques are presented in the discussion, This section discusses the comparison of conventional machine learning techniques and deep learning, available MRI datasets, MRI modalities, MRI preprocessing toolboxes, DL models, DL toolboxes, and classifier methods.

The most important challenges of MS diagnosis with MRI modalities and DL techniques are delineated. The inaccessibility of huge sMRI datasets belonging to diverse population and lack of access to fMRI modalities are among the most important dataset-related challenges which are discussed in detail. Moreover, DL-related challenges include researchers’ lack of access to powerful hardware resources for MS diagnosis research.

Future work suggestions are presented in a section of the paper. They focus on developing more available public datasets of sMRI modalities, functional neuroimaging modalities (fMRI and PET), and implementation of rehabilitation systems for MS patients.

Appendix A. Details of deep learning architectures for MS researches

| Works | DNN | Details | Classifier | Loss Function | Optimizer |
|---|---|---|---|---|---|
| Marzullo [149] | 2D-CNN | 2 Conv + 2 Max Pooling + 2 Dropout + 2 BN + 2 FC | Linear | -- | Adam |
| Siar [154] | 2D-CNN | 25 Layers | Softmax | -- | -- |
| Aslani [155] | 2D-CNN | ResNet50 + UFF Blocks | -- | BCE | Adadelta |
| Eitel [158] | 2D-CNN | 5 Conv + 5 BN + PIF | -- | -- | -- |
| Afzal [159] | 2D-CNN | 2 Conv + 2 Max Pooling + 1 FC | Multinomial LR | -- | -- |
| Roy [172] | 2D-CNN | 15 Conv | -- | -- | Adam |
| Aslani [184] | 2D-CNN | 3 Parallel ResNet50s + 5 MMFF Blocks + 4 MSFU Blocks + MPR Block | Softmax | soft Dice Loss function | Adam |
| Alijamaat [190] | 2D-CNN | 15 Conv + 1 Average Pooling + 1 FC + Dropout | Sigmoid | -- | Adam |
| Shrwan [223] | 2D-CNN | 3 Conv + 3 BN + 3 Max Pooling + 2 FC | Softmax | CE | SGDM |
| Afzal [225] | Two 2D-CNN | 6 Conv + 6 Max Pooling | -- | -- | Proposed |
| Wang [181] | 2D-CNN | 11 Conv + 11 BN + 4 Pooling + 3 FC + 2 Dropout | Softmax | -- | -- |
| Ulloa [135] | V-Net CNN | 3 Conv + 3 Max Pooling + 2 FC + 5 Dropout | Sigmoid | BCE and Focal Loss | SGD |
| Birenbaum [179] | 4 CNN Models | V-Net, L-Net, Multi-View CNN, Multi-View Longitudinal CNN | Softmax | CCE | Adadelta |
| Birenbaum [165] | 4 CNN Models | V-Net, L-Net, Multi-View Longitudinal CNN, Multi-View CNN | -- | CCE | Adadelta |
| SALEM [156] | Encoder-Decoder U-NET | 2 Encoders and 2 Decoders | -- | CCE | Adadelta |
| | Cascaded 3D CNNs | Cascade of 2 Identical CNNs | | | |
| Roca [127] | 3D-CNN | 6 Conv + 3 BN + 3 Max Pooling + 2 Dense | Linear Activation | MSE | Adam |
| Nair [129] | 3D-CNN | 12 Conv + 4 De-Conv + 4 Dropout +4 Skip Connection | Sigmoid | Weighted BCE | Adam |
| Brown [132] | 3D-CNN | 6 Conv + 4 De-Conv and Up Sampling + 4 Concatenation | Softmax | CCE | Adam |
| Sepahvand [152] | 3D CNN | 10 Conv + 4 Max Pooling + 4 BN + 4 Dropout + 2 FC | Sigmoid | CE | Adam |
| | Modified U-net | 17 Conv + 7 BN + Dropout + 3 Max Pooling + 3 De-Conv + 3 Concatenation | | | |
| Rosa [153] | Cascade of Two 3D Patch-Wise CNNs | 4 Conv + 2 Max Pooling + 4 BN + 1 FC + 1 Dropout | Softmax | CE | Adam |
| Tousignant [162] | 3D-CNN | 3 Consecutive Conv Blocks + 2 FC + 5 Dropout | Sigmoid | CE | RMSProp |
| Yoo [163] | 3D-CNN | 3 Conv + 3 Max Pooling +2 FC + 2 Dropout | LR | CE | Adadelta |
| Kazancli [168] | Two 3D-CNNs in a Cascade Fashion | 2 Conv + 2 Average Pooling + 2 BN + 1 FC + 1 Dropout | Softmax | CE | Adam |
| Gros [169] | Sequence of Two CNNs | First CNN with 2D Dilated Convolutions, Second CNN with 3D Convolutions | -- | Dice Loss | Adam |
| Valverde [173] | 3D-CNN | 4 Conv + 2 Max-Pooling + 4 BN + 3 FC + 3 Dropout | Softmax | CCE | Adadelta |
| Zhang [176] | 3D-CNN | 7 Conv +7 Pooling + 3 FC + 3 Dropout | Softmax | -- | -- |
| Yoo [178] | 3D-CNN | 3 Conv + 3 Max Pooling + 2 FC + 2 Dropout | LR | CE | Adadelta |
| Eitel [188] | 3D-CNN | 4 Conv + 4 Max-Pooling + 4 Dropout | Sigmoid | -- | Adam |
| Valverde [192] | 3D-CNN | 2 Conv + 2 Pooling + 1 Dropout + 1 FC | Softmax | CCE | Adadelta |
| Valverde [175] | Cascade of Two 3D Patch-Wise CNNs | 2 Conv + 2 Max Pooling + 2 BN + 1 FC + 1 Dropout | Softmax | CCE | Adadelta |
| Gessert [170] | Attention-Guided Two-Path CNNs | 2 Conv In + 21 ResBlocks + 6 Conv Down + 3 Conv Up + Fusion Block + Conv Out | -- | Dice Loss | Adam |
| Sepahvand [116] | NE SubNet | 17 Conv + 7 BN + 3 Max Pooling + 3 De-Conv + 4 Concatenation | Sigmoid | CE | Adam |
| McKinley [121] | DeepSCAN | 2 Conv Blocks + 1 Max Pooling block + 4 Dilated Dense Blocks + 1 Up Sampling | -- | Combination of Focal Loss and Label-Flip Loss | -- |
| Ackaouy [122] | Seg-JDOT | 6 Conv + 5 Context Modules + 4 Up Sampling + 3 Localization Modules | Softmax | Proposed | Proposed |
| Maggi [128] | CVSnet | 3 Conv + 3 Max Pooling + 3 Dropout + FC | Softmax | CCE | Adam |
| McKinley [227] | DeepSCAN | 2 Conv + 4 Dense Blocks + Max Pooling + Up Sampling | -- | Combination of Multi-Class CE Loss and Label-Flip Loss | Adam |
| HASHEMI [189] | 3D Patch-Wise FC-Dense-Net | 5 Conv + 3 BN + 11 DenseBlocks + 5 Transition Down + 5 Transition Up + 1 De-Conv + 5 Concatenation | Sigmoid | Asymmetric Loss Functions | Adam |
| McKinley [171] | DeepSCAN | Cascade of Two CNNs | Softmax | Hybrid Loss | Adam |
| Vincent [114] | FiLMed-Unet | -- | -- | Dice Loss | Adam |
| Vang [130] | Synergy-Net | Fusing U-Net and Mask R-CNN and RPN Sub-Networks | -- | Multi-Tasks Loss Function | Adam |
| Calimeri [193] | Graph Based Neural Networks | Vertex Sequential Fully Connected (vs-FC) + the Graph Sequential Fully Connected (gs-FC) + Dropout | Softmax | -- | Adamax |
| Marzullo [186] | Graph Convolutional Neural Network (GCNN) | 1 Graph Conv + FC + Dropout | Softmax | -- | Adam |
| Dai [187] | MDN | Cascading 2 Basic Blocks (Dilated Convolutions, Global and Local Residual Learnings, Concatenation Layers) | -- | Proposed Loss Function | Adam |
| Dewey [183] | DeepHarmony | 10 Conv + 8 Strided Conv + 17 BN + 5 Concatenation | -- | MAE | Adam |
| Yoo [161] | Hierarchical Multimodal Fusion (HMF) Model | 3 Conv + 3 Max Pooling + 6 FC + 3 RBM + 2 mf-fc + 1 hf-fc + 6 Dropout | Logistic Regression | CE | AdaDelta |
| Essa [134] | 2 Parallel R-CNN | 6 Conv + 3 Polling + 2 FC + Softmax | ANFIS | -- | -- |
| Hou [148] | Cross Attention Densely-Connected Network (CA-DCN) | 3 Cross Attention Block + 12 Conv + 3 Down Sampling + 3 Up Sampling +8 Concatenation | 3 Softmax | Proposed | -- |
| Ulloa [150] | Single-View Multi-Channel (SVMC) | 3 Conv + 3 Max Pooling + 4 Dropout + 1 FC | Softmax | CCE | SGD |
| Zhang [151] | Recurrent Slice-Wise Attention Network (RSANet) | 3D U-Net Backbone with RSA Blocks | -- | Exponentially Weighted CE | Adam |
| Narayana [117] | VGG16+FCN | Modified Architecture + 3 FC | Sigmoid | BCE | Adam |
| Barquero [133] | RimNet (two parallel CNNs inspired by VGGNet) | 12 Conv + 6 Max Pooling + 3 BN + 3 FC | Softmax | CE | Adam |
| Ye [140] | DNN | 10 Hidden FC + 10 BN | Softmax | CE | Adam |
| Fenneteau [167] | 3D U-Net | 26 Conv + 4 Strided-2 Conv + 30 Instance Normalization + 5 Dropout + 7 Addition + 6 Up-Sampling + 4 Concatenation | Sigmoid | -- | Adam |
| Coronado [119] | 3D U-Net | 5 Conv + 4 Context Modules + 3 Up Sampling Modules + 2 Localization Modules + 2 Segmentation + 3 Strides + 3 De-Conv + 1 Upscaling | Softmax | Multiclass Weighted Dice | Adam |
| Narayana [120] | Multiclass U-net | 18 Conv + 4 Max Pooling + 4 De-Conv and Up Sampling + 4 Copy and Concatenation | Softmax | Balanced Version of Dice Score Coefficient | SGD |
| Rosa [123] | Multi-task 3D U-Net + ICD | 9 Conv + 2 Max Pooling + 2 Up Sampling + 3 Concatenation | -- | Voxel-Wise Weighted CE | Adam |
| Rosa [124] | 3D U-Net | 7 Conv + 2 Max Pooling + 2 De-Conv + 2 Concatenation | -- | pixel-wise weighted CE | Adam |
| Narayana [126] | 2D U-net | 16 Conv + 4 Max Pooling + 4 De-Conv and Up Sampling + 4 Copy and Concatenation | Softmax | Balanced Version of Dice Score Coefficient | Adam |
| Abolvardi [141] | 3D U-Net | 19 Conv + 4 Max Pooling + 4 Up Sampling and Conv + 4 Copy and Crop | -- | -- | -- |
| Falvo [142] | Multimodal Dense U-Net (MDU) | 11 Conv + 3 Pooling + 1 Merge and Conv + 2 De-Conv + 6 Dense Blocks + 3 Copy and Concatenation | -- | Proposed | Adam |
| Ghosal [143] | Light-Weighted U-Net | 10 Conv + 8 BN + 4 Max Pooling + 4 Up Sampling | Sigmoid | BCE | Adam |
| Kumar [144] | Modified Dense U-Net | 6 Dense Blocks + 3 Max Pooling + 5 Conv + 3 Up Sampling + 3 Concatenation | Softmax | BCE | Adam |
| Kats [146] | 2D U-net based FCNN | 6 Conv + 2 Max Pooling + 2 Dropout + 2 De-Conv + 2 Concatenation | Sigmoid | Proposed | -- |
| Feng [147] | 3D U-Net | 15 Conv + 14 BN + 3 Max Pooling + 3 De-Conv and Up Sampling + 3 Copy and Crop | -- | Weighted CE | Adam |
| Narayana [157] | Multi-Class U-Net | 18 Conv + 4 Max Pooling + 3 De-Conv and Up Sampling + 4 Copy and Concatenation | -- | Weighted CCE | Adam |
| Hu [160] | 3D Attention Context U-Net (ACU-Net) | 2 Conv + 5 3D Context Guided Modules + 2 3D Spatial Attention Blocks + 3 De-Conv + 3 Channel-Wise Concatenation | Softmax | Focal Tversky Loss function | SGD |
| Gabr [164] | Multiclass U-Net FCNN | 18 Conv + 4 Max Pooling + 4 De-Conv and Up Sampling + 4 Copy and Concatenation | -- | Multiclass Dice Loss | Adam |
| Salem [137] | FCNN | 3D Registration Architecture + 3D Segmentation Architecture | -- | Proposed | Adam |
| Yoo [177] | Multimodal Deep Learning Network | 2 DBNs | RF | -- | -- |
| Sujit [145] | Ensemble DL Model | 3 Cascaded Networks (Each Cascaded Network Consists of a DCNN Followed by a Fully Connected Network | Averaging the Quality Scores | BCE | Adam |
| Finck [115] | DiamondGAN | 2 Generators, 2 Discriminators | 2 Neuroradiologists | Cycle Consistency Loss Function | -- |
| Wei [125] | Conditional Flexible Self-Attention GAN (two CF-SAGAN used as Sketcher and Refiner) | Generator:2 Conv + 4 ResDown Blocks + 2 Flexible Self-Attention + 4 ResUp Blocks + 4 Long Connections<br>Discriminator: Conv + 4 ResDown Blocks + 1 Flexible Self-Attention + 1 Dense | Sigmoid | Adversarial Loss Functions | Adam |
| Shaul [138] | GAN | 2 Generator (2 U-Nets), 1 Discriminator | Sigmoid | Proposed | Adam |
| Zhang [180] | MS-GAN | Multimodal Encoder-Decoder Generator + Multiple Discriminators | -- | Proposed | Adam |
| Wei [182] | Sketcher-Refiner GANs | 2 cGANs named Sketcher and Refiner | Softmax | Adversarial Loss Functions | Adam |
| Wei [185] | Sketcher-Refiner GANs | 2 cGANs named Sketcher and Refiner | Softmax | Adversarial Loss | Adam |
| Hagiwara [224] | Conditional GAN | Generator: 2 Parallel Fully Connected Neural Network Streamlines<br>Discriminator: Similar to The Structure Of U-Net | Sigmoid | Adversarial Loss | Adam |
| Karaca [166] | SSAE | 2 Autoencoders | Softmax | Proposed | -- |
| Vogelsanger [226] | Introspective Variational Autoencoder (intro-VAE or IVAE) | Encoder: Conv + BN + Pooling + Dense + Dropout<br>Decoder: Dense + Dropout + BN + De-Conv + Up Sampling | LDA | Proposed | -- |
| Aslani [118] | Traditional Encoder-Decoder Network with Regularization Network | Encoder Network + Decoder Network + Regularization Network | Softmax | Proposed | Adam |
| McKinley [191] | Nabla-Net | 17 Conv + 16 BN + 3 Max Pooling + 3UnPooling + Concatenation | Sigmoid | BCE | Adadelta |
| Krüger [136] | Fully 3D Convolutional Encoder-Decoder Architecture | 37 Conv + 5 Up Sampling and Conv + 9 Concatenation + 12 Element Wise Sum + 3 Segmentation + 2 Up Scale | -- | CE | Adam |
| Brosch [174] | Convolutional Encoder Network with Shortcut Connections (CEN-s) | 2 Conv + 1 Average Pooling + 2 De-Conv + 1 UnPooling | -- | Proposed | Adadelta |
| Tripathi [139] | CNN-DMRI | 3 Conv + 2 Down Sampling + 4 Residual Blocks + 2 De-Conv | -- | MSE | Adam |
| Gessert [113] | Enc-convGRU-Dec | 2 Conv + 12 ResBlock + 3 Conv Down + 3 Conv Up + 4 convGRU | -- | -- | -- |
| Andermatt [131] | MD-GRU | -- | Softmax | -- | -- |
| Sander [194] | MD-GRU | -- | -- | -- | -- |